\documentclass[journal=jctcce,manuscript=article]{achemso}

\usepackage[version=3]{mhchem} % Formula subscripts using \ce{}
\usepackage{upgreek}
\usepackage{physics}
\usepackage{amsmath,amsfonts,amsthm,amssymb,bm}
\usepackage[
	colorlinks=true,
    citecolor=blue,
    linkcolor=blue,
    filecolor=blue,      
    urlcolor=blue,	
    breaklinks=true,
	unicode=true, psdextra]{hyperref}

\usepackage{subcaption}
\usepackage{color}
\usepackage{comment}
\usepackage{siunitx}
\usepackage{multirow}
\usepackage{rotating} % For the sidewaystable environment
\DeclareSIUnit\angstrom{\text {Å}}
\newcommand{\ddv}[1]{{\color{blue}#1}}

\usepackage{xcolor}
\definecolor{purple}{HTML}{993bff}

\newcommand{\ricechem}{Department of Chemistry, Rice University, Houston, Texas 77005-1892, USA}
\newcommand{\ricephys}{Department of Physics and Astronomy, Rice University, Houston, TX 77005-1892, USA}
\newcommand{\ucl}{Department of Chemistry, University College London, 20 Gordon Street, London, WC1H 0AJ, U.K.}

\author{Don~Danilov}
\affiliation{\ricechem}
\author{Hugh~G.~A.~Burton}
\affiliation{\ucl}
\author{James~Shee}
\affiliation{\ricechem}
\alsoaffiliation{\ricephys}
\email{james.shee@rice.edu}

\title[CSF-ROHF ph-AFQMC]
  {Open-shell Configuration State Functions as trial states for phaseless auxiliary-field quantum Monte Carlo}

\begin{document}

% \begin{tocentry}

% Some journals require a graphical entry for the Table of Contents.
% This should be laid out ``print ready'' so that the sizing of the

% Inside the \texttt{tocentry} environment, the font used is Helvetica
% 8\,pt, as required by \emph{Journal of the American Chemical

% The surrounding frame is 9\,cm by 3.5\,cm, which is the maximum
% permitted for  \emph{Journal of the American Chemical Society}
% graphical table of content entries. The box will not resize if the
% content is too big: instead it will overflow the edge of the box.

% This box and the associated title will always be printed on a
% separate page at the end of the document.

% \end{tocentry}

\begin{abstract}
    The accurate description of open-shell states in molecular systems poses challenges for approximate electronic structure methods. 
   One emerging approach is to variationally optimize the orbitals of a single user-selected Configuration State Function (CSF), i.e., a linear combination of Slater determinants that is constructed as an eigenstate of the total spin operator. In this work we explore the use of these orbital-optimized single CSFs as trial wavefunctions in phaseless auxiliary-field quantum Monte Carlo (ph-AFQMC), a stochastic method that recovers electron correlation by evolving an initial state in imaginary time, subject to a constraint imposed by the trial state. We investigate the symmetric stretch of \ce{H-He-H}, a series of small $p$-block compounds with diradicaloid singlet states, twisted ethene conformers, and a bimetallic iron-sulfur complex.  We find that accurate energetics are obtained for states dominated by open-shell character, and that for certain atoms and diatomics the accuracy can be significantly improved when the trial is made to also preserve the relevant angular momentum symmetries.  Furthermore, our study reveals that the open-shell CSF trial is suboptimal when the target electronic state reflects a significant mixture of ionic and covalent character, e.g., at shorter bond lengths in \ce{H-He-H} and for the superexchange-stabilized ground state of [2Fe--2S]$^{2+}$.    
    Our findings suggest that ph-AFQMC using CSF trial states is a promising method for many (but 
    not all) strongly correlated molecular states, especially given its relatively low computational cost, and that it provides a reasonable non-perturbative way to recover some of the dynamic correlation that is otherwise missing in the CSF. 
    
\end{abstract}

\section{Introduction}

Molecular electronic states can be classified as closed-shell, where all spatial orbitals are doubly occupied, or open-shell, where a subset of spatial orbitals is singly occupied.
In reality, many chemical ground states fall somewhere along the continuum between these two limits.  %While have (potentially low lying) open shell excited states, systems where the ground state is open shell are relatively uncommon, typically arising from some orbital degeneracy. Note that nuclear distortions can also introduce open shell character, especially when the distortion introduces orbital degeneracies (for example, stretching H$_2$ to where the $\sigma$/$\sigma*$ orbitals become nearly degenerate). 
Understanding the behavior of predominately open-shell states is fundamental to describing a wide range of chemical phenomena including photochemical pathways\cite{vincentThatLittleExtra2016, zhangNonadiabaticExcitedStateMolecular2020}, catalysis,\cite{nandyWhyConventionalDesign2020, parkMechanismRhCatalyzedOxidative2016} and exotic spin materials\cite{yuanFractionalSpinonQuasiparticles2025, broholmQuantumSpinLiquids2020, aliQuantumSpinSensors2025, dongRecentProgressOpenshell2022, tanElectronicSpintronicOpenShell2022, tasiuSpintronicsTechnologyComprehensive2026}.

Modeling open-shell states poses distinct challenges for electronic structure theory, many of which arise  %describing a mean field reference (necessary for many post-Hartree Fock methods including Coupled Cluster and Møller–Plesset perturbation theory) that can be radically different between up and down electrons and the 
due to their inherently multi-configurational nature.\cite{Krylov2017}  While spin unrestricted formalisms %allowing for the system to have different molecular orbitals for the up and down electrons helps address the first issue, 
can yield qualitatively correct dissociation curves even at the mean-field level, this advantage comes at the cost of breaking physical spatial or spin symmetries\cite{Lykos1963} (e.g., $\hat{S}^2$),  which can complicate a subsequent perturbation theory treatment. %in an unrestricted Hartree Fock (UHF) calculation.  %the second challenge inherently limits the potential application of single-configurational methods. 
Multi-configurational methods, such as Complete Active Space Self Consistent Field (CASSCF),\cite{lengsfieldGeneralSecondOrder1980,Roos1980a,Roos1980b, eadeDirectMinimizationMCSCF1981, roosMulticonfigurationalQuantumChemistry2016} %require a trade-off of including enough configurations to properly describe the electronic correlation and the increase in associated computational cost; 
are viable for systems where a relatively small active space can be clearly defined, but the exponential scaling of such methods with active space size is computationally prohibitive for larger systems. Methods like Multi-Reference Configuration Interaction (MRCI)\cite{yamaguchiMRCI1980, plasserMultiReferenceConfigurationInteraction2020} and various types of selected CI\cite{Sharma2017, kossoskiHierarchyConfigurationInteraction2022} can reach slightly larger system sizes with the caveat that these methods are susceptible to missing families of important configurations requiring careful user-selection of starting determinants.

These challenges raise a natural question: what if instead of attempting to \emph{discover} the multi-configurational structure of a wavefunction by solving for the relevant CI coefficients, we \emph{posit} a fixed determinant expansion based on a known spin coupling pattern % (as the analytical forms for any possible open shell coupling are trivial to calculate) 
and optimize the orbitals for this constrained wavefunction? 
Such an approach can be motivated by the fact that many low-spin antiferromagnetic states are known to be dominated by a particular spin-coupled Configuration State Function (CSF) in a localized orbital basis,\cite{Iszak2023,Manni2020,Dobrautz2022,MartiDafcik2024}
compressing the amount of variational information in the wavefunction and increasing the 
Hamiltonian sparsity when represented in the space of CSFs.
The orbital optimization for such a wavefunction has been termed spin-adapted Configuration State Function restricted open-shell Hartree--Fock (CSF-ROHF) and recent implementation advances allow for efficient optimization of CSF molecular orbital coefficients for large systems such as a 15-fused polyacene\cite{Burton2025} and multi-metal center organometallic clusters.\cite{Gouveia2024}
However, numerical results have shown that %documented a number of instances where chemical predictions from 
orbital-optimized CSFs are unable to correctly predict low-spin ground states 
in compounds such as sulfur-bridged di-iron clusters due to the lack of 
charge transfer configurations in the CSF-ROHF wavefunction.\cite{Gouveia2024,Song2026}   

One motivation of the present work is to explore a low-scaling avenue to recover the  %(mostly ``dynamic'') 
electron correlation missing from a single CSF-ROHF wavefunction 
by using this approximation as a trial state for phaseless auxiliary-field quantum Monte Carlo (ph-AFQMC).  ph-AFQMC is a non-perturbative method that stochastically estimates the ground-state energy along an imaginary-time propagation in the space of Slater determinants (\textit{vide infra}).  
However, like other quantum Monte Carlo methods, AFQMC suffers from the fermionic sign problem\cite{beccaQuantumMonteCarlo2017, lohSignProblemNumerical1990} which requires the introduction of the phaseless constraint.
This constraint can give rise to biases that make the imaginary-time trajectory and resulting accuracy sensitive to the choice of trial wavefunction -- an effect most acutely felt in strongly correlated systems.
Many types of trial states have been proposed and assessed in the literature for open-shell systems, including various types of symmetry-broken Hartree--Fock solutions\cite{purwantoEliminatingSpinContamination2008,leeUtilizingEssentialSymmetry2020, sheeSingletTripletEnergy2019,danilovCapturingStrongCorrelation2025}, CI subspace-projected Coupled Cluster\cite{mahajanCCSDTAccuracyLower2025,kjonstad2025systematic,kjonstad2026can}, CASSCF\cite{leeTwentyYearsAuxiliaryField2022,sheePotentiallyTransformativeRole2023}, selected CI\cite{mahajanSelectedConfigurationInteraction2022,neugebauerBenchmarkQualityInitioPredictions2023,vuong2026evaluating}, matrix product states\cite{jiangUnbiasingFermionicAuxiliaryfield2025}, non-orthogonal CI\cite{landinezbordaNonorthogonalMultiSlaterDeterminant2019, sukurmaSelfRefinementAuxiliaryFieldQuantum2025},  stochastically-sampled variational trials,\cite{xiaoImplementingAdvancedTrial2025} neural-network quantum states,\cite{xiao2025nnqs} and quantum-computer derived trials\cite{danilovEnhancingAccuracyEfficiency2025, rofougaranBenchmarkingQuantumTrial2026}. Whether some aspects of ph-AFQMC can be offloaded to quantum devices is also an ongoing area of research\cite{hugginsUnbiasingFermionicQuantum2022, zhaoQuantumClassicalAuxiliaryField2025, huangEvaluatingQuantumclassicalQuantum2024}.  

Despite a growing body of work on the ph-AFQMC methodology, fundamental questions remain open about the role of the trial wavefunction and the requisite properties of an optimal one, especially for strongly correlated systems.  For example,  recent work using spin symmetry-broken mean-field trials on active space models of synthetic iron-sulfur clusters has revealed that such trials can lead to a sampling bias (in addition to the phaseless bias) that can result in overly negative ph-AFQMC energies.\cite{danilovSelectingOptimalUnrestricted2026}  Furthermore, correlating the symmetry-broken mean-field states with a hierarchy of truncated cluster operators did not lead to systematically improved energies, %versus exact or near-exact reference values, 
and sometimes led to ph-AFQMC energies that were more positive than the employed trial states.\cite{kjonstad2026can}  It is reasonable to hypothesize that %correlated \emph{spin-adapted} 
trial wavefunctions that capture spin-correlation absent in a broken-symmetry mean-field description could lead to more accurate ph-AFQMC energies for strongly correlated ground states that are dominated by open-shell character, which motivates the present study of CSF-ROHF trial states.

This work is structured as follows: after a brief overview of the relevant theory, we will assess the performance of CSF-ROHF trials in ph-AFQMC calculations for a series of open-shell systems.  Beginning with the symmetric dissociation of \ce{H-He-H}, an exactly-solvable model system exhibiting tunable superexchange physics, we will then explore a set of $p$-block organics with triplet ground states and diradicaloid singlet excited states.  Moving to larger systems, we will consider ethene (ethylene) twisted about its dihedral angle by 82.5 and 90 degrees, and finally a dinuclear iron-sulfur compound in both active space and full orbital space contexts.  Our assessment of CSF-ROHF trials paints a practical picture of when these trials are expected to perform well, e.g., for open-shell states of purely diradical character in molecules of `low' point-group symmetry.  Our study also reveals shortcomings and findings that are of fundamental interest to the ph-AFQMC methodology, including the importance of %using trials that preserve 
$L_z$ symmetry conservation in the trial for certain atomic and linear molecules, and of capturing ionic contributions in the trial wavefunction for systems such as iron-sulfur clusters.

\section{Theoretical background}
\label{sec:theory}

\subsection{Genealogical Configuration State Functions}

A configuration state function (CSF) is defined as a wavefunction that is an eigenfunction of 
both the total spin operator $\hat{S}^2$ and the spatial orbital occupation number operators 
%\ddv{(in second quantization)}
$\hat{n}_p = \sum_{\sigma \in \{\uparrow,\downarrow \}} \hat{a}_{p\sigma}^{\dagger}\hat{a}_{p\sigma}^{\vphantom{\dagger}}$.\cite{helgakerMolecularElectronicstructureTheory2000}
This formalism encodes a particular orbital configuration while retaining spin symmetry.
A single CSF requires a linear combination of Slater determinants to provide a spin 
eigenfunction for an open-shell configuration, providing a generalization of a closed-shell 
Slater determinant for unpaired electrons.
The coefficients of this linear combination can be derived analytically using the genealogical 
coupling scheme, where electrons are coupled sequentially using the Clebsch--Gordan coefficients
such that the wavefunction remains an eigenstate of $\hat{S}^2$ at every step.
The spin structure of a CSF is encoded in its genealogical spin-coupling pattern that describes
how the total spin changes as each electron is coupled.
For example, the open-shell spin coupling pattern $++--$ represents two electrons coupled to a triplet, 
and then a further two electrons coupled to give an overall singlet.
Crucially, unlike a single Slater determinant, the wavefunction is no longer invariant to rotations that mix occupied orbitals.\cite{Burton2025}
We direct the interested reader to Ref.~\citenum{Manni2021} for a detailed overview of this construction.
Genealogical CSFs have a long history of use in spin-pure electronic structure theory
due to the ease of computing matrix elements through the graphical unitary group approach.
\cite{Paldus1974, Paldus1976a, shavittGraphicalUnitaryGroup1981}

It has been recently recognized that certain low-spin open-shell ground states can be 
qualitatively described by a single CSF using a suitably chosen set of molecular 
orbitals.\cite{Dobrautz2022,Manni2021a,Manni2023,MartiDafcik2025,Iszak2023,MartiDafcik2024}
By switching from a delocalized to a localized orbital picture, many low-spin antiferromagnetic states that 
have no dominant Slater determinant can be well-represented by a single CSF. 
This perspective relies on a suitable spin coupling pattern and ordering of localized orbitals 
that reflects the physical coupling in the system.
For example, the asymptotic ground state for \ce{N2} dissociation can be described
by a single CSF with the open-shell spin-coupling pattern $+++---$ 
(representing two quartets localized on each \ce{N} coupled into an overall singlet) in contrast to 400 active space determinants.\cite{MartiDafcik2024} 
It has been shown that this representation can increase the sparsity of the Hamiltonian in the CSF basis and
improve the convergence of Full Configuration Interaction quantum Monte Carlo,\cite{Dobrautz2019,Manni2020,Dobrautz2022} 
while also allowing systematic targeting of ground and excited states.\cite{Manni2021b}

A natural extension is to use a single CSF with an arbitrary spin-coupling pattern
as the target wavefunction in a self-consistent-field (SCF) calculation, enabling ROHF calculations
for low-spin open-shell states.
This approach has historically been challenging due to the lack of a robust SCF theory for 
arbitrary open-shell configurations, which arises because different orbitals experience a 
different Fock operator.\cite{Roothaan1960}
Such calculations have only recently become possible through the derivation of new algorithms 
based on Fock diagonalization for an arbitrary spin-coupling pattern,\cite{Gouveia2024} quasi-Newton optimization,\cite{Burton2025}  and the super-CI formalism.\cite{Song2026}
Here, we employ the CSF-based extension of geometric direct minimization,\cite{Burton2025} which 
relies on analytic energies and gradients that can be computed at mean-field cost.
Full details of the optimization algorithm can be found in Ref.~\citenum{Burton2025}.

While these algorithms enable the optimization of the molecular orbitals for 
an arbitrary genealogical CSF, they have also confirmed that a single CSF is not always sufficient 
to predict low-spin antiferromagnetic ground states due to the absence of ionic charge-transfer
configurations.\cite{Gouveia2024,Burton2025,Song2026}
This finding poses the question of whether the remaining electron correlation required for low-spin 
stabilization can be captured from ph-AFQMC using a CSF-ROHF trial, providing one motivation for the current work.

\subsection{Phaseless auxiliary-field quantum Monte Carlo}

We consider a non-relativistic Born--Oppenheimer molecular electronic Hamiltonian in standard second quantization form: 
\begin{equation}
    \hat{H} = \sum_{pq, \sigma}h_{pq}\hat{a}^\dagger_{p\sigma} \hat{a}_{q\sigma} +\sum_{pqrs,\sigma\tau} V_{pqrs} \hat{a}^\dagger_{p\sigma} \hat{a}^\dagger_{q\tau} \hat{a}_{s\tau} \hat{a}_{r\sigma} = \hat{H}_1 + \hat{H}_2.
\end{equation}
We can group the one electron terms into $\hat{H}_1$ and the two electron terms in $\hat{H}_2$. Given some initial wavefunction $|\Psi_i\rangle$ that has non-zero overlap with the exact target eigenstate, AFQMC can purify the ground state via imaginary-time propagation, i.e., \begin{equation}
    \lim_{\tau\rightarrow\infty} e^{-\tau (\hat{H} - E_0)}|\Psi_i\rangle = \lim_{P\rightarrow\infty} \big[e^{-\Delta\tau (\hat{H} - E_0)}]^P|\Psi_i\rangle = |\Psi_0\rangle, 
\end{equation}
where the propagation can be broken up into $P$ steps of size $\Delta \tau = \tau / P$. Note that all excited states will decay away under imaginary time evolution. 

By invoking a pre-computed trial state $\Psi_T$, the wavefunction is expressed in importance-sampled form as a weighted sum of non-orthogonal determinants as
\begin{equation}
    \Psi \sim \sum_k \frac{w_k|\phi_k\rangle}{\langle \Psi_T|\phi_k\rangle}.
\end{equation}
The $k$-th walker is then characterized by a weight $w_k$, a Slater determinant $\ket{\phi_k}$, and an overlap with the trial state $\langle \Psi_T|\phi_k\rangle$.
At each imaginary-time step, the one- and two-body parts of the Hamiltonian can be split with a Trotter--Suzuki decomposition\cite{suzukiGeneralizedTrottersFormula1976, trotterProductSemigroupsOperators1959}:
\begin{equation}
    e^{-\Delta\tau\hat{H}} = e^{-\Delta\tau\hat{H_1}/2}e^{-\Delta\tau\hat{H_2}}e^{-\Delta\tau\hat{H_1}/2} + \mathcal{O}(\Delta\tau^3)
\end{equation}
such that Thouless' theorem\cite{thoulessStabilityConditionsNuclear1960} can be invoked to realize the application of a one-body propagator as an orbital rotation, i.e., $e^{-\Delta\tau\hat{H_1}}|\phi\rangle = |\phi'\rangle$.
The two-body term can always be decomposed into a sum of squares of one-body operators, $\hat{H_2} = -\frac{1}{2}\sum_\alpha \hat{v}^2_\alpha$, and the Hubbard--Stratonovich transformation\cite{hubbardCalculationPartitionFunctions1959, stratonovichMethodCalculatingQuantum1957} is used to decouple the interacting system into a sum of non-interacting systems, each in an auxiliary-field as
\begin{equation}
    e^{\frac{1}{2}\Delta\tau\hat{v}^2_\alpha} = \int_{-\infty}^\infty dx_\alpha P(x_\alpha) e^{\sqrt{\Delta\tau}x_\alpha\hat{v}_\alpha}, 
\end{equation}
where $P(x_\alpha)$ is a Gaussian function.  
This enables the propagation of each walker determinant by a product of three one-body operators, $
    \hat{B}(\mathbf{x}) = e^{-\Delta\tau\hat{H}_1/2} e^{\sqrt{\Delta\tau}\mathbf{x}\cdot\mathbf{\hat{v}}} e^{-\Delta\tau\hat{H}_1/2},
$ while the weight update involves multiplication by $\frac{\langle \Psi_T | \hat{B}(\mathbf{x})|\phi\rangle}{\langle \Psi_T|\phi\rangle}$.   The above procedure is exact, and the energy can be estimated from averaging the following over $\tau$, $E_{\text{MC}}(\tau) = \frac{\sum_k w_k(\tau)E_{\text{L}}^k(\tau)}{\sum_k w_k(\tau)}$, where the `local energy' is defined as $E_L^k = \frac{\langle \Psi_T|\hat{H}|\phi_k\rangle}{\langle \Psi_T|\phi_k\rangle}$.
However, due to the fermionic sign problem there is an exponentially decaying signal to noise ratio (though subtraction of $\langle \hat{\mathbf{v}}\rangle_{\Psi_T}$ from $\hat{\mathbf{v}}$ in the propagator helps to some extent with numerical stability\cite{haoAuxiliaryFieldQuantum2019}), requiring additional constraints that can lead to an uncontrolled bias.

The phaseless constraint is discussed in detail elsewhere \cite{zhangQuantumMonteCarlo2003, shiSymmetryAuxiliaryfieldQuantum2013, mottaInitioComputationsMolecular2018, danilovSelectingOptimalUnrestricted2026} and therefore will only be summarized here.  Shifting the integration variable by a `force-bias' chosen to remove any accumulated phase to first order in $\Delta\tau$ enables the weight update: 
$w(\tau+\Delta\tau) = w(\tau) e^{-\Delta\tau E_\text{L}}$.  After each propagation step, the real part of $E_\text{L}$ is kept, and walkers that have phase rotations in excess of $\pm \frac{\pi}{2}$ are removed (and later replaced at regular intervals by larger-weighted walkers).  The walker weights then remain real and positive, and determinant sign-changes are prevented.  In the limit of an exact trial wavefunction, the averaged phaseless AFQMC energy will always be exact, though the stochastic wavefunction may not be\cite{liFokkerPlanckEquationGoverning2026}.  

\section{Results}

\subsection{H-He-H symmetric stretch}

\begin{figure}
  \centering
  \includegraphics[width=.32\linewidth]{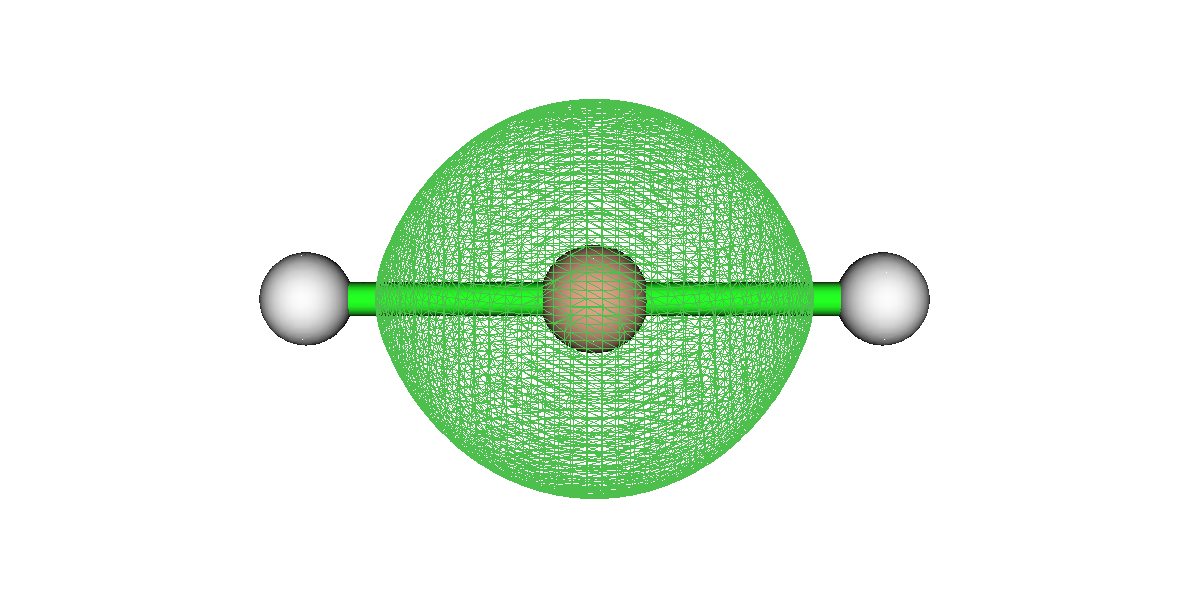}
  \includegraphics[width=.32\linewidth]{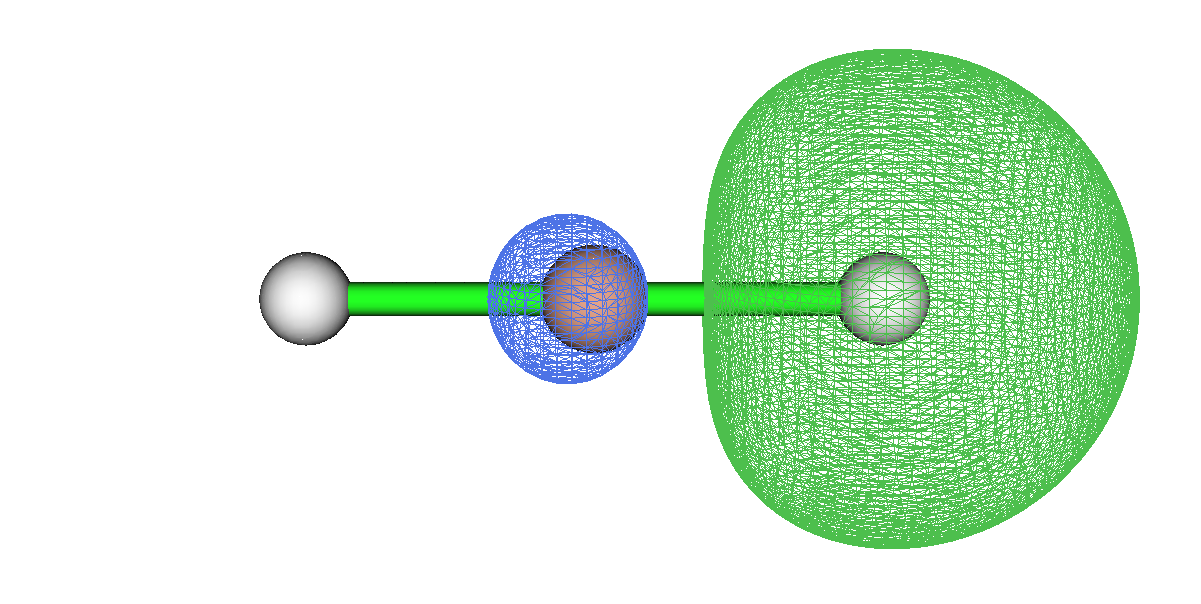}
  \includegraphics[width=.32\linewidth]{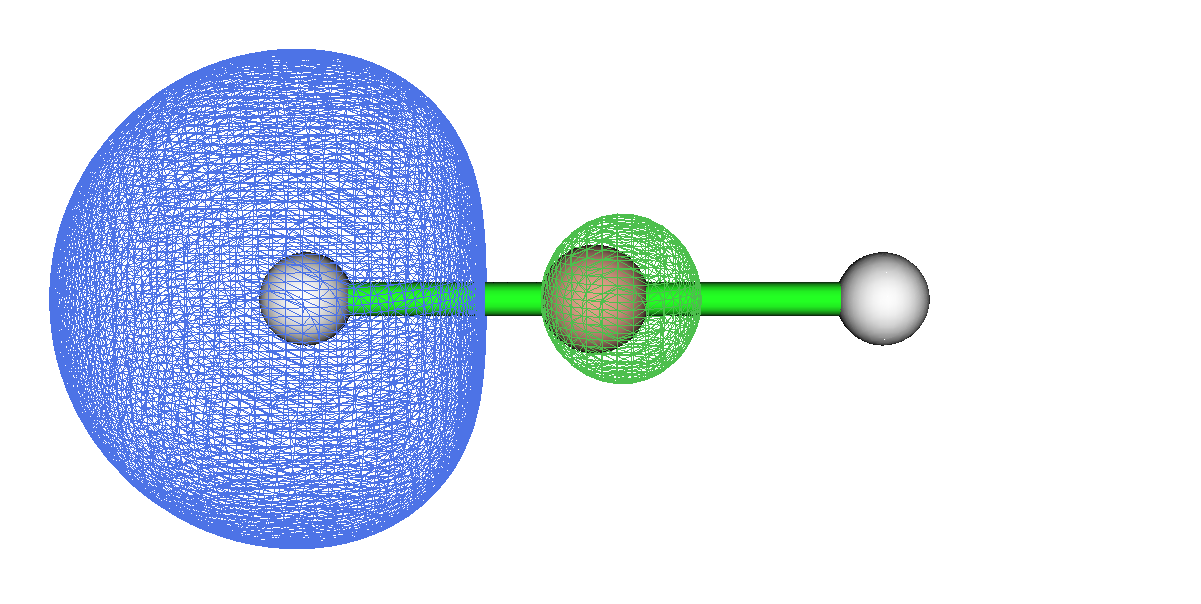}

  \caption{\footnotesize CSF-ROHF occupied molecular orbitals for HHeH at an \ce{H-He} bond distance of \SI{1.5}{\angstrom}. The first orbital (left) is dominated by the He 1s orbital and is doubly-occupied in the CSF-ROHF wavefunction, whereas the other two orbitals (middle, right) are singly occupied and are predominantly localized on the two hydrogens.}
  \label{fig:hheh:mos}
\end{figure}

The ground-state singlet of the linear \ce{H-He-H} molecule, along its totally symmetric ($\Sigma_\text{g}^+$) stretch, is a simple strongly correlated system with tunable open-shell character. In a localized basis at stretched bond lengths, one finds roughly two spin-coupled unpaired electrons on the two H atoms (\autoref{fig:hheh:mos}).  This open-shell singlet state has previously been studied computationally as a minimal magnetic model complex that can capture, to some extent, superexchange effects.\cite{schurkusTheoreticalPredictionMagnetic2020,liuPracticableParameterDescribing2001,weiLocalSpinStudy2005} Further motivating its inclusion in this study is the trivial ability to control the ratio of open- to closed-shell character by symmetrically stretching the \ce{H-He} bonds, and the ease of obtaining exact reference energies (we use the cc-pVDZ basis set\cite{Dunning1989, woon1993a}).

\begin{figure}[ht]
  \centering
  \begin{subfigure}[b]{0.49\textwidth}
    \centering
    \includegraphics[width=\linewidth]{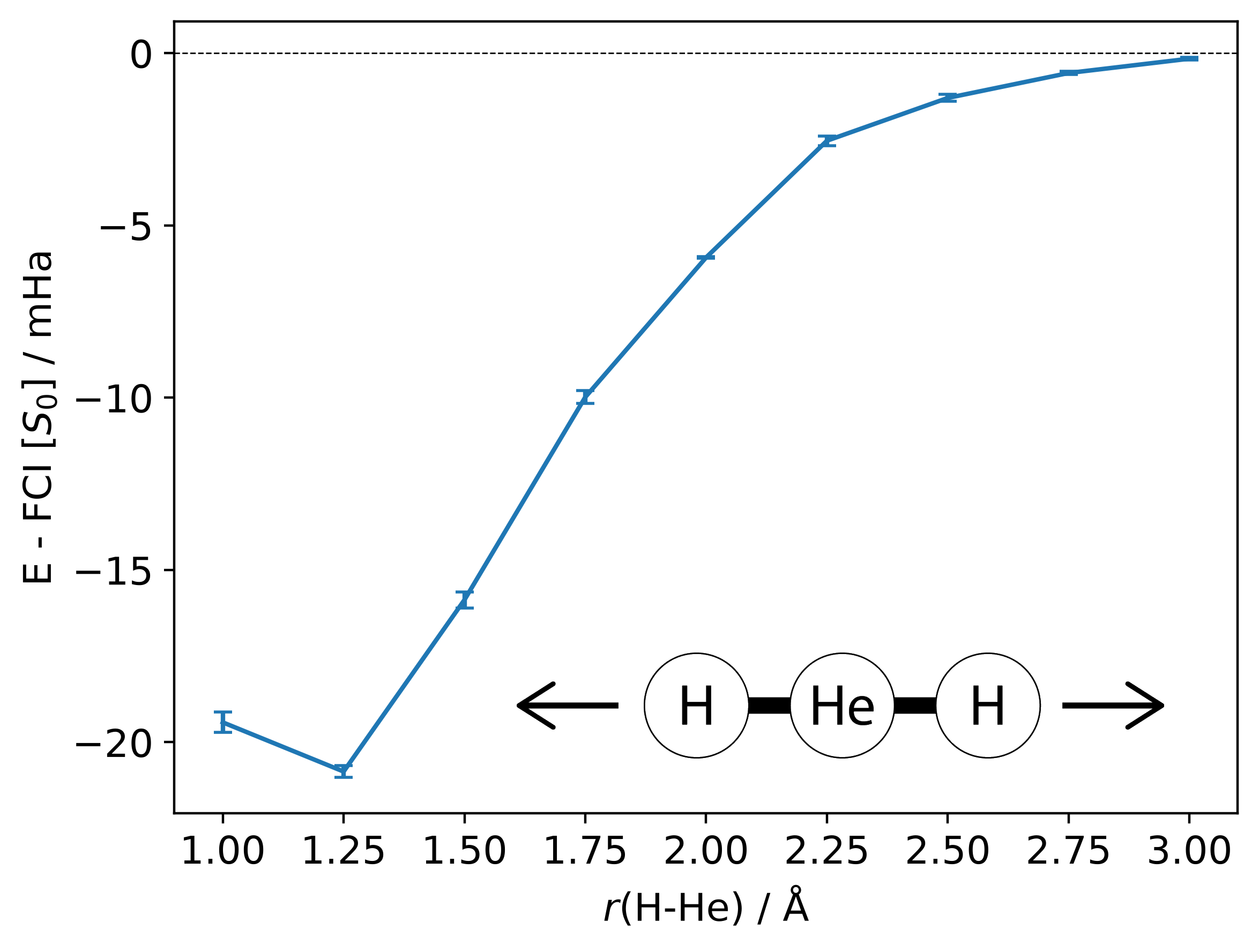}
    \caption{}
  \end{subfigure}
  \hspace*{\fill}
  \begin{subfigure}[b]{0.49\textwidth}
    \centering
    \includegraphics[width=\linewidth]{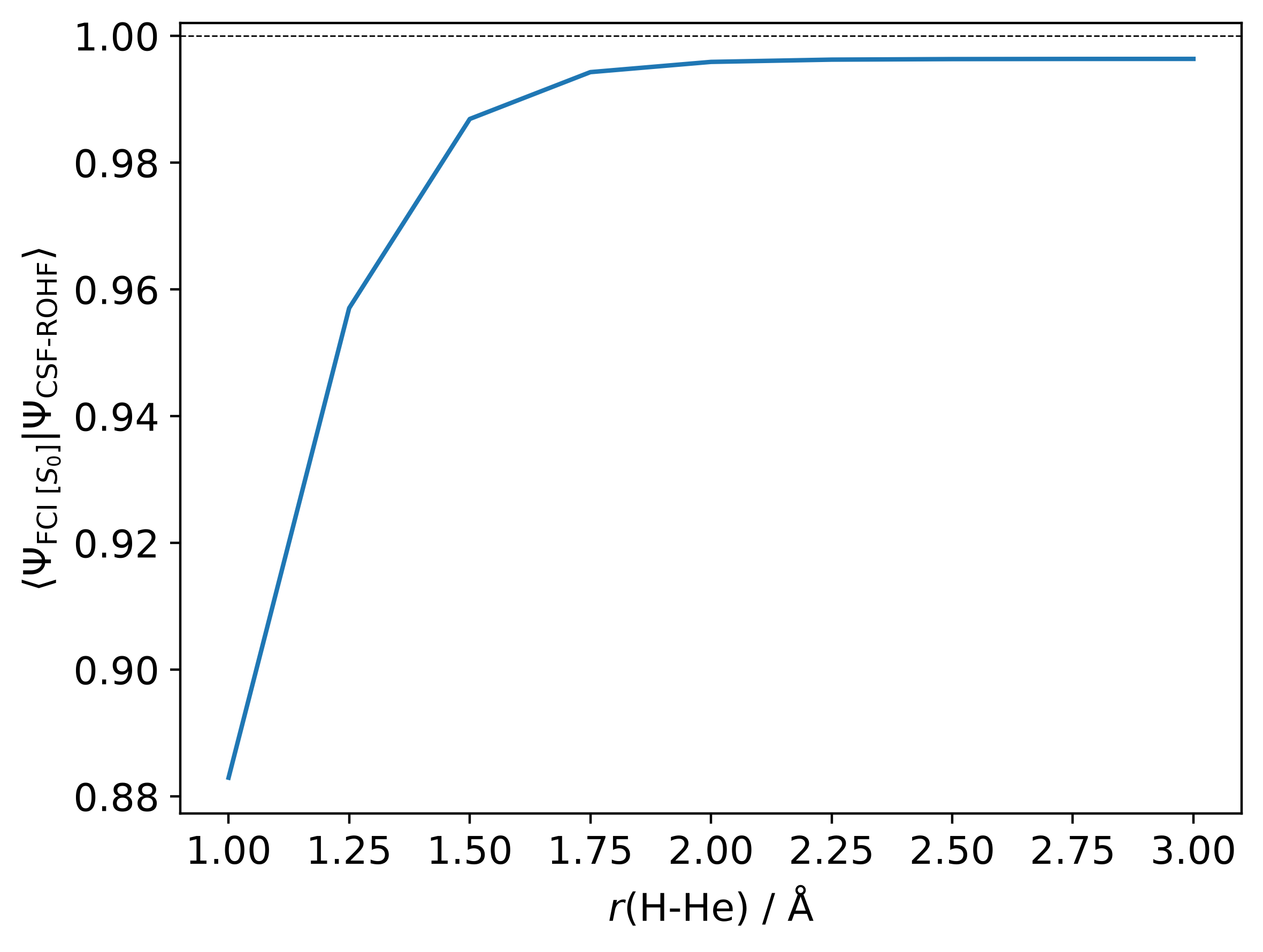}
    \caption{}
  \end{subfigure}
  \caption{\footnotesize (a) ph-AFQMC energies (relative to FCI) using CSF-ROHF trial states and (b) the overlaps of CSF-ROHF trial wavefunctions with the exact singlet ground states, for the symmetric $\Sigma_\text{g}^+$ stretch in \ce{HHeH} at different \ce{H-He} bond lengths.
  }
  \label{fig:hheh:csf1}
\end{figure}

Along the $\Sigma_\text{g}^+$ stretching coordinate, ph-AFQMC with CSF-ROHF trial wavefunctions using the $+-$ spin coupling yields systematically lower energies relative to the exact values, with the underestimation generally becoming less severe at longer distances (\autoref{fig:hheh:csf1}).  
While ph-AFQMC is well documented to be non-variational\cite{taheridehkordiPhaselessAuxiliaryField2023, carlsonIssuesObservationsApplications1999, leePerformancePhaselessAuxiliaryfield2020,danilovSelectingOptimalUnrestricted2026}, this result is somewhat surprising given the reasonably high overlap of the CSF-ROHF trial states with the FCI states across the dissociation coordinate ($88\%$ at the shortest bond length $r(\text{\ce{H-He}}) = \SI{1}{\angstrom}$).  As this overlap approaches unity at longer bond lengths, the ph-AFQMC energy deviations from FCI decrease and eventually converge towards the exact result, reflecting the increase in ground-state open-shell character towards the dissociation limit.  Note that the overlap never reaches \emph{exactly} one, even at large \ce{H-He} distances, as the exact wavefunction has small contributions from configurations involving excitations from the doubly-occupied core orbital (at longer distances this is dominated by the Helium $1s$), which is strictly doubly-occupied in the CSF-ROHF wavefunction by construction. 

\begin{figure}[ht]
  \centering
  \includegraphics[width=0.6\linewidth]{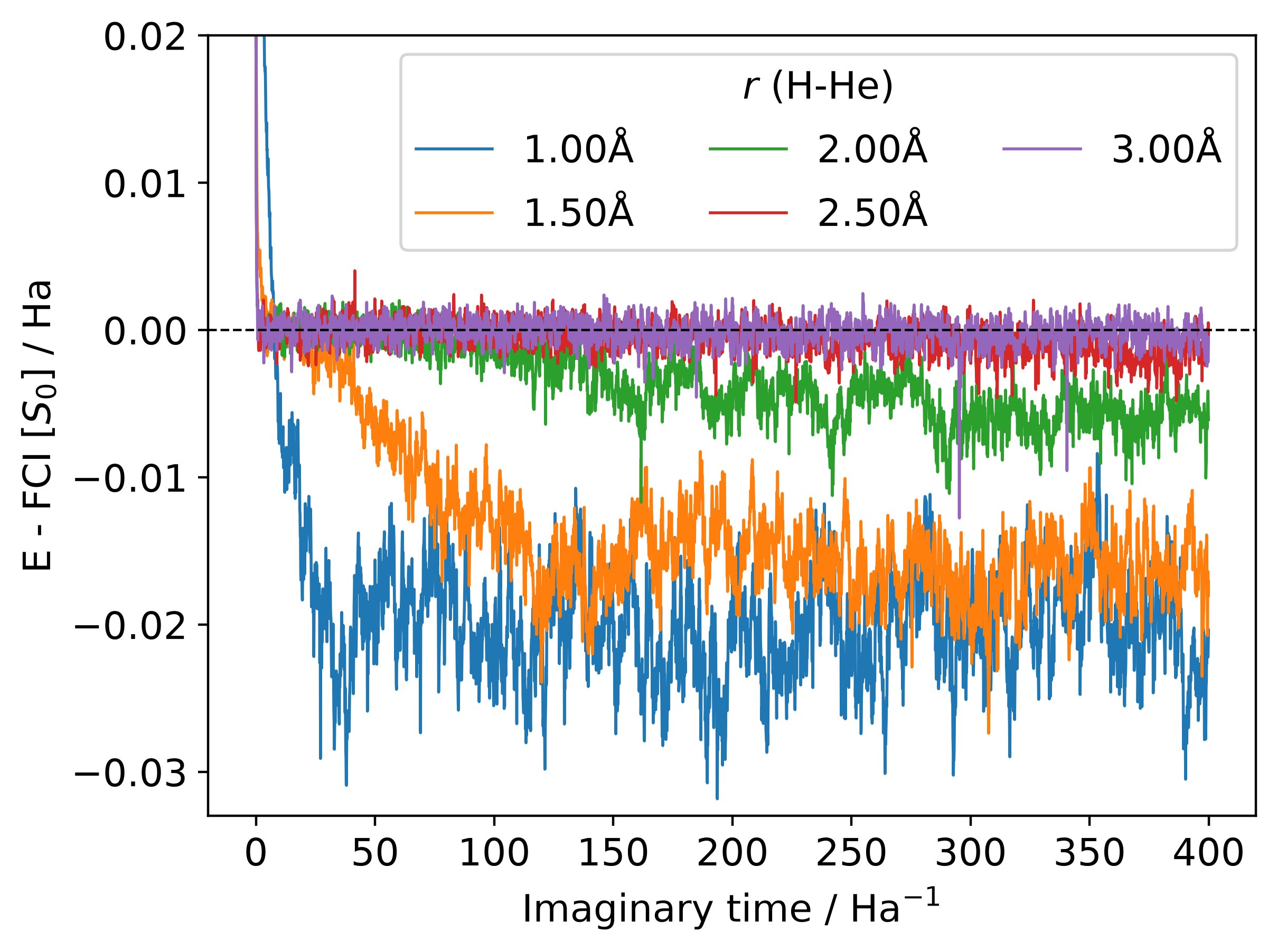}
  \caption{\footnotesize Representative imaginary time ph-AFQMC trajectories for the $\Sigma_g^+$ stretch of the HHeH molecule at five different distances using a CSF-ROHF trial, plotted relative to respective FCI ground state energies. Note the systematic underestimation of the exact energy that becomes less pronounced at longer distances, and also the `shoulders' at early imaginary times near the FCI energy for 1.5\AA{} and 2.0\AA{} trajectories.}
%  \hb{[HB: Can we cut the y-axis to zoom in on the important region and show the difference between trajectories more clearly? Also the legend text is a bit small.]}\ddv{[Since the initial energy is not important to the story here I made a  zoomed in plot w/ larger font labels]}}
  \label{fig:hheh:csf2}
\end{figure}

\begin{figure}[ht]
  \centering
\includegraphics[width=\linewidth]{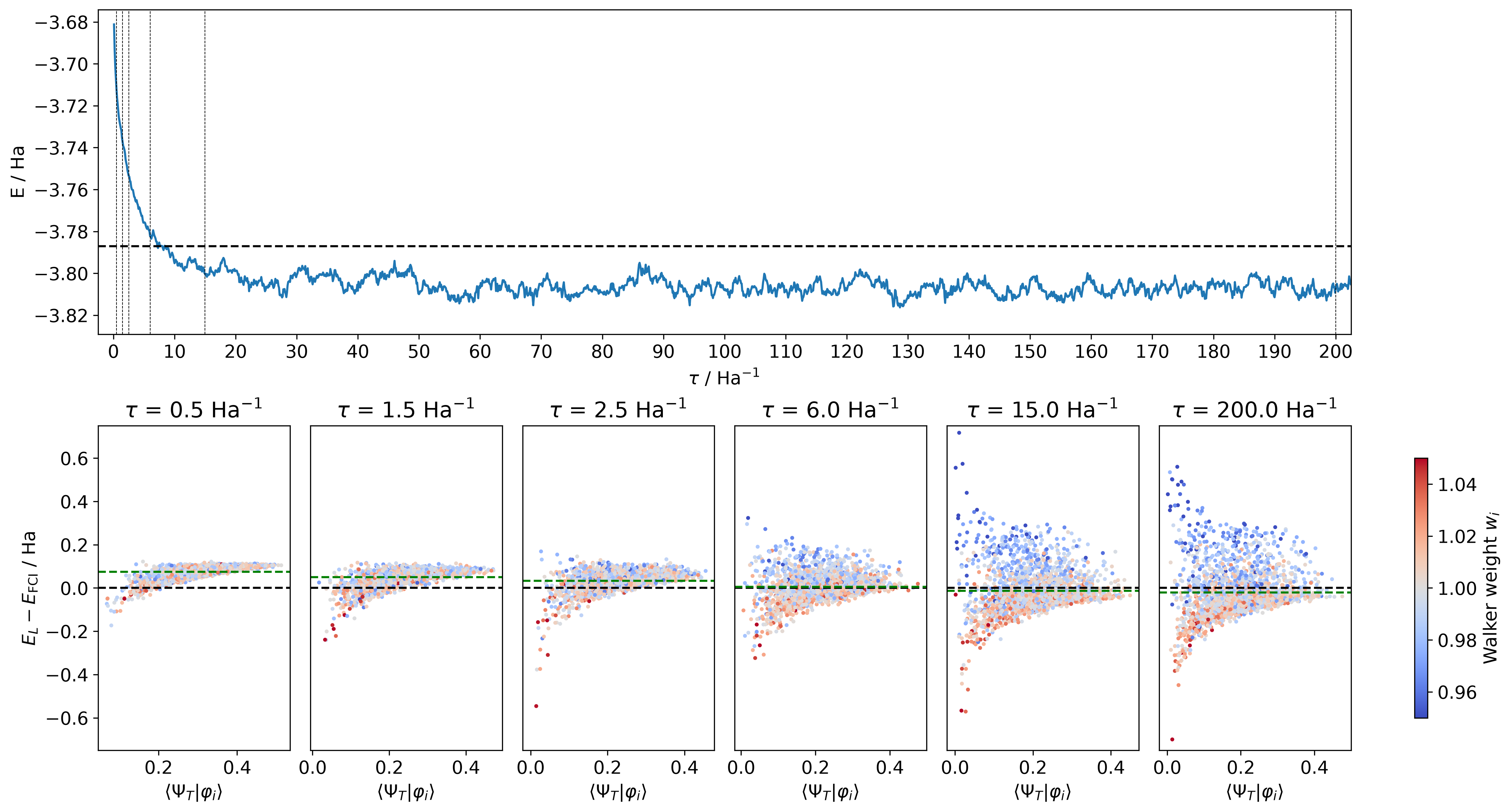}
  \caption{\footnotesize (top) Imaginary time trajectory for a CSF-ROHF trial ph-AFQMC calculation for HHeH at $r(\text{\ce{H-He}})=\SI{1}{\angstrom}$ (a zoomed in version of the relevant trajectory shown in \autoref{fig:hheh:csf2}). (bottom) The distribution of walker energies (relative to FCI shown as a black dashed horizontal line) and the QMC energy averaged over walkers (green dashed line) at different imaginary times (marked in dashed vertical lines on upper plot) with the dot color indicating the walker weights. Note the introduction of comparatively heavily-weighted, yet spuriously low energy walkers as the imaginary time propagation proceeds, leading to below-FCI ph-AFQMC energies.}
  \label{fig:hheh:elocal_olap_weight}
\end{figure}

\autoref{fig:hheh:csf2} shows the energy of imaginary-time trajectories corresponding to ph-AFQMC calculations with CSF-ROHF trial states at various \ce{H-He} distances.  The trajectories at shorter \ce{H-He} distances have a noticeable shoulder at early imaginary times that consistently occurs near the FCI energy, before proceeding to overly negative energies.  
%To investigate this shoulder, 
The dynamic evolution of the walker population is monitored at the representative short bond length $r$(\ce{H-He}) = \SI{1}{\angstrom}. 
At various imaginary-time slices, \autoref{fig:hheh:elocal_olap_weight} shows:  (i) the overlap of the walker determinants with the trial wavefunction; (ii) the deviation of the local energies from the exact energy; and (iii) the walker weights.  
At early times, the walkers have relatively uniform weights with an average energy (shown by the dashed green line) that is higher than the FCI result. After roughly 6 Ha$^{-1}$, the average energy starts to dip below the exact value, and evidently walkers with overly negative local energies tend to have larger weights (exceeding unity).  This observed behaviour is consistent with the previously hypothesized mechanism that gives rise to a sampling bias, which results when the trial wavefunction fails to favor walkers that have high overlap with itself.\cite{danilovSelectingOptimalUnrestricted2026}  That is,  when a walker determinant has small overlap with the trial state, the local energies can be spuriously negative, which tends to increase the weight of that walker, leading eventually to overly negative ph-AFQMC energy estimates.  We observe that this bias plays out in a relatively gradual manner (as opposed to a single numerical explosion), likely because of the rare-event bounds designed to cap the weights and force-bias magnitudes\cite{purwanto2009pressure}.  These bounds vanish in the limit of $\Delta\tau \rightarrow 0$.  This result is a first indication that open-shell CSF-ROHF trial wavefunctions can be suboptimal when the nature of the target electronic state deviates from pure open-shell character.

\begin{figure}[htb]
  \centering
  \includegraphics[width=0.48\linewidth]{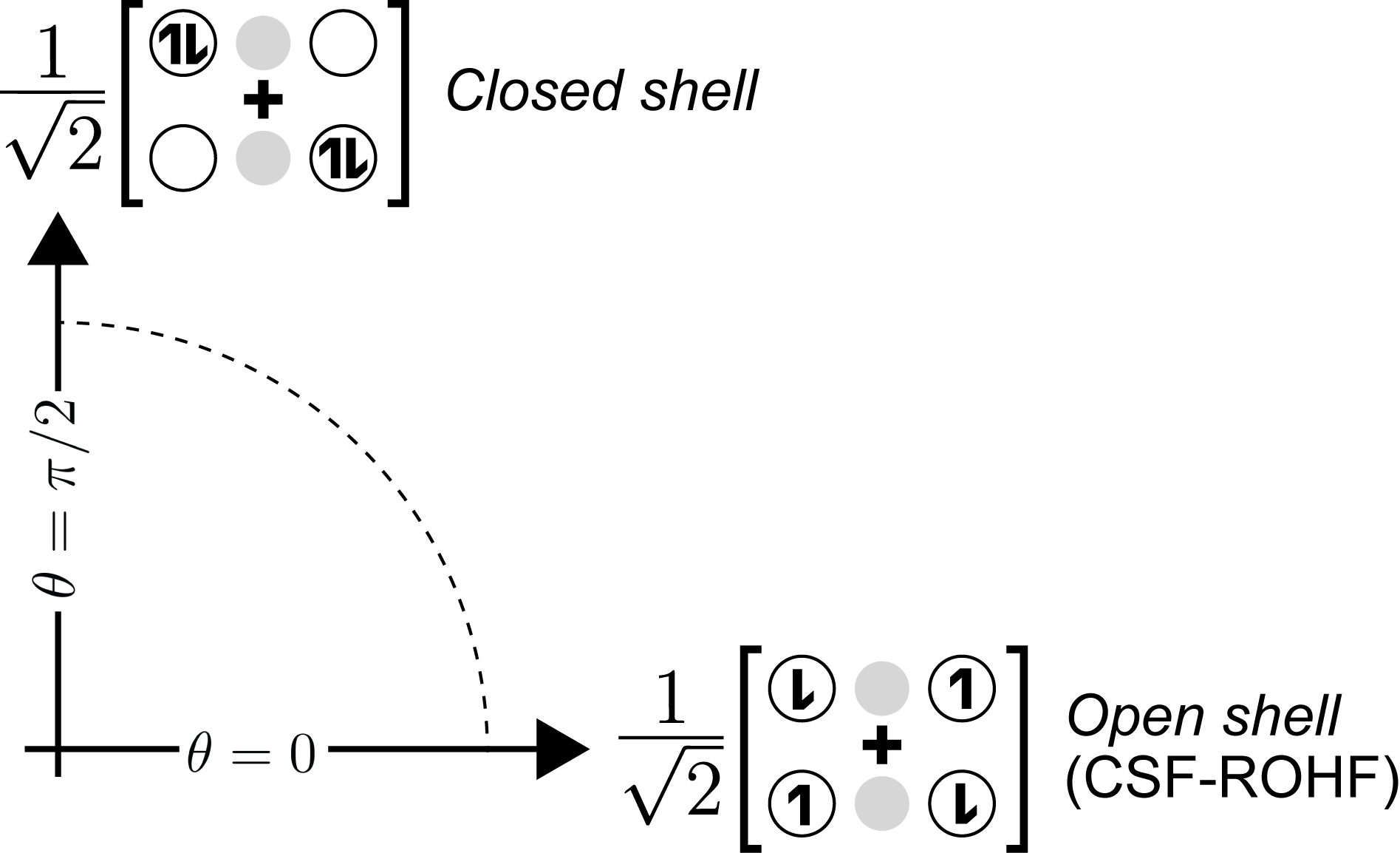}
  \caption{\footnotesize Schematic showing the wavefunction dependence on the mixing 
  parameter $\theta$  in Eq.~\eqref{eq:mixing}.}
  \label{fig:hheh:theta_cartoon}
\end{figure}

To gain more insight into how the nature of the trial wavefunction can affect the ph-AFQMC imaginary-time trajectory and the accuracy of the energy estimate,  %into the role of the trial wavefunction's overlap with the FCI ground state on the accuracy of resulting ph-AFQMC calculations beyond a single point at each nuclear geometry, 
we consider a fixed bond length of HHeH and a trial wavefunction that depends on a single continuous parameter $\theta$, which defines a mixing angle between open-shell and closed-shell character:
\begin{equation}
    \ket{\Psi_T(\theta)} = \frac{\sin\theta}{\sqrt{2}}( \ket{\text{220}} + \ket{\text{202}} ) - \frac{\cos\theta}{\sqrt{2}} ( \ket{2\hspace*{-2pt}\uparrow \downarrow} + \ket{2\hspace*{-2pt}\downarrow \uparrow} )
    \label{eq:mixing}
\end{equation}
where the occupancies of the three orbitals (shown in \autoref{fig:hheh:mos}) are denoted by the following convention:  $0$ corresponds to no electrons, $\uparrow$ to a single spin-up electron, $\downarrow$ to a single spin-down electron, and $2$ to double-occupancy.  As discussed above, the first orbital (dominated by the He 1s core orbital) is \emph{always} doubly occupied in this model. %and the $\frac{1}{\sqrt{2}}$ factors arise from normalization. %$\braket{\Psi_T(\theta)} = 1$. 
$\ket{\Psi_T(\theta = 0)}$ corresponds to the idealized open-shell CSF-ROHF wavefunction discussed above (up to a phase);  $\ket{\Psi_T(\theta = \pi/2)}$ corresponds to an equal mixture of closed-shell determinants that, at larger distances, is an excited singlet state. 
A schematic of the state and its dependence on mixing angle is shown in \autoref{fig:hheh:theta_cartoon}. 

\begin{figure}[htb!]
  \centering
  \begin{subfigure}[b]{0.48\textwidth}
    \centering
    \includegraphics[width=\linewidth]{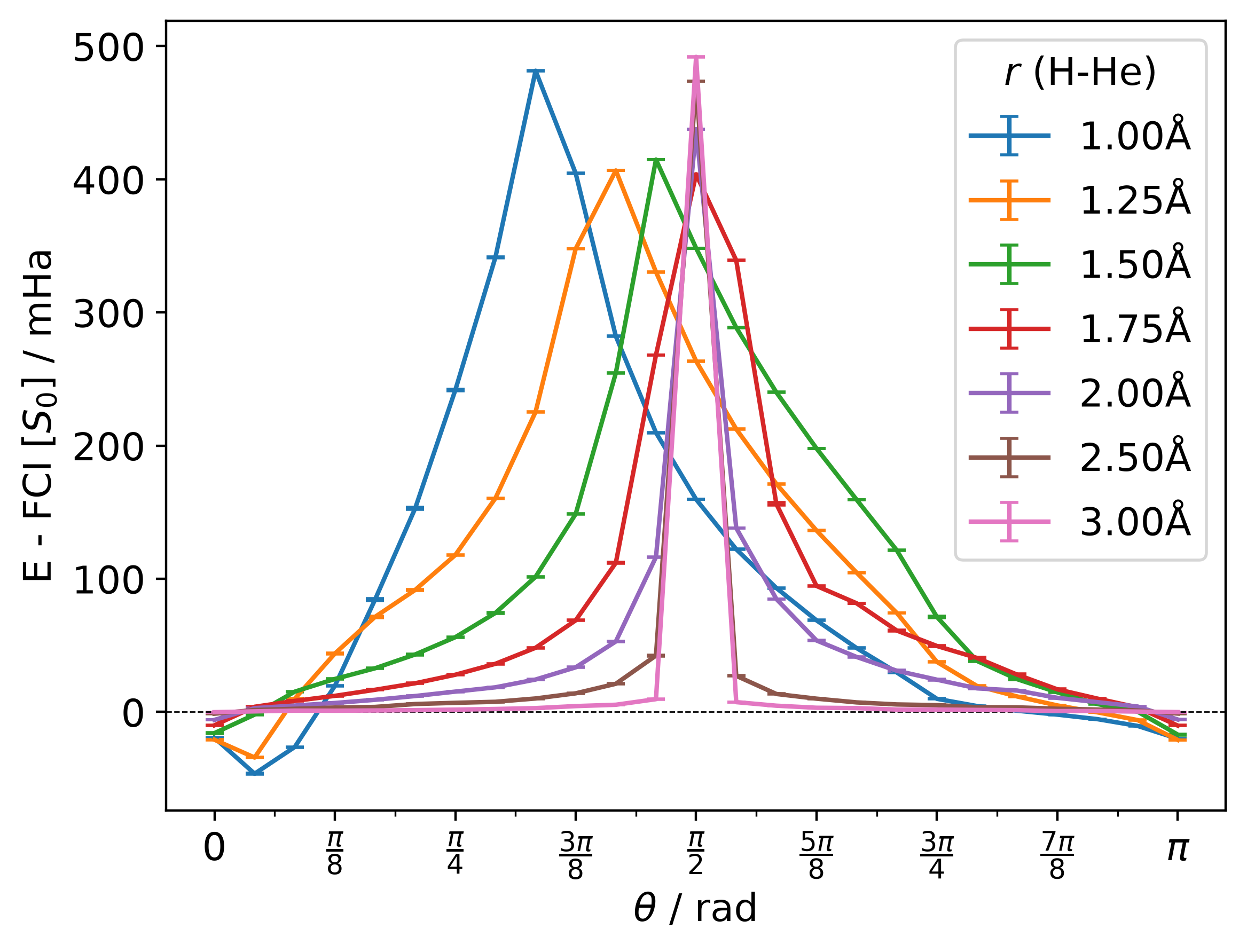}
    \caption{}
  \end{subfigure}
  \begin{subfigure}[b]{0.48\textwidth}
    \centering
    \includegraphics[width=\linewidth]{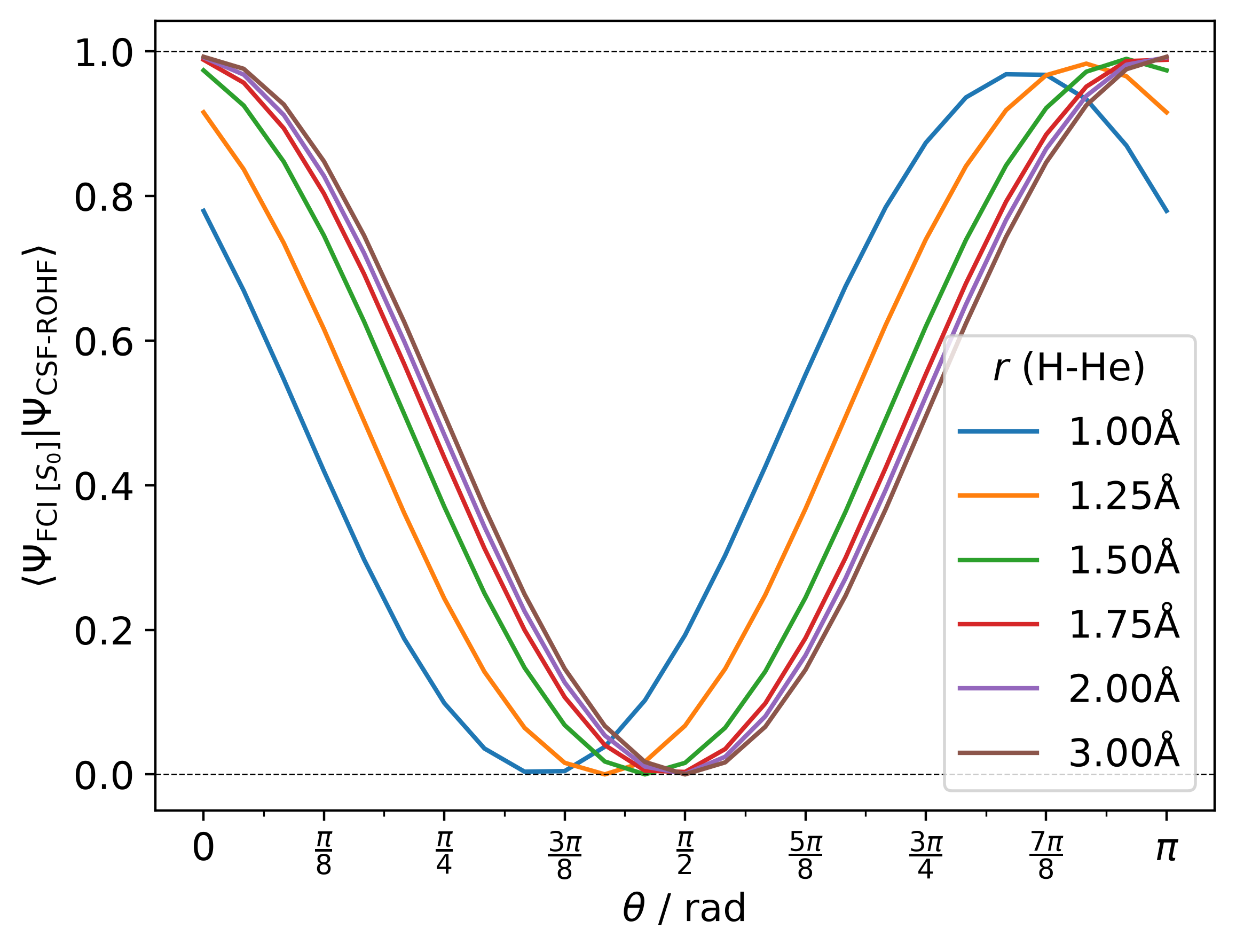}
    \caption{}
  \end{subfigure}
  \\
  \begin{subfigure}[b]{0.48\textwidth}
    \centering
    \includegraphics[width=\linewidth]{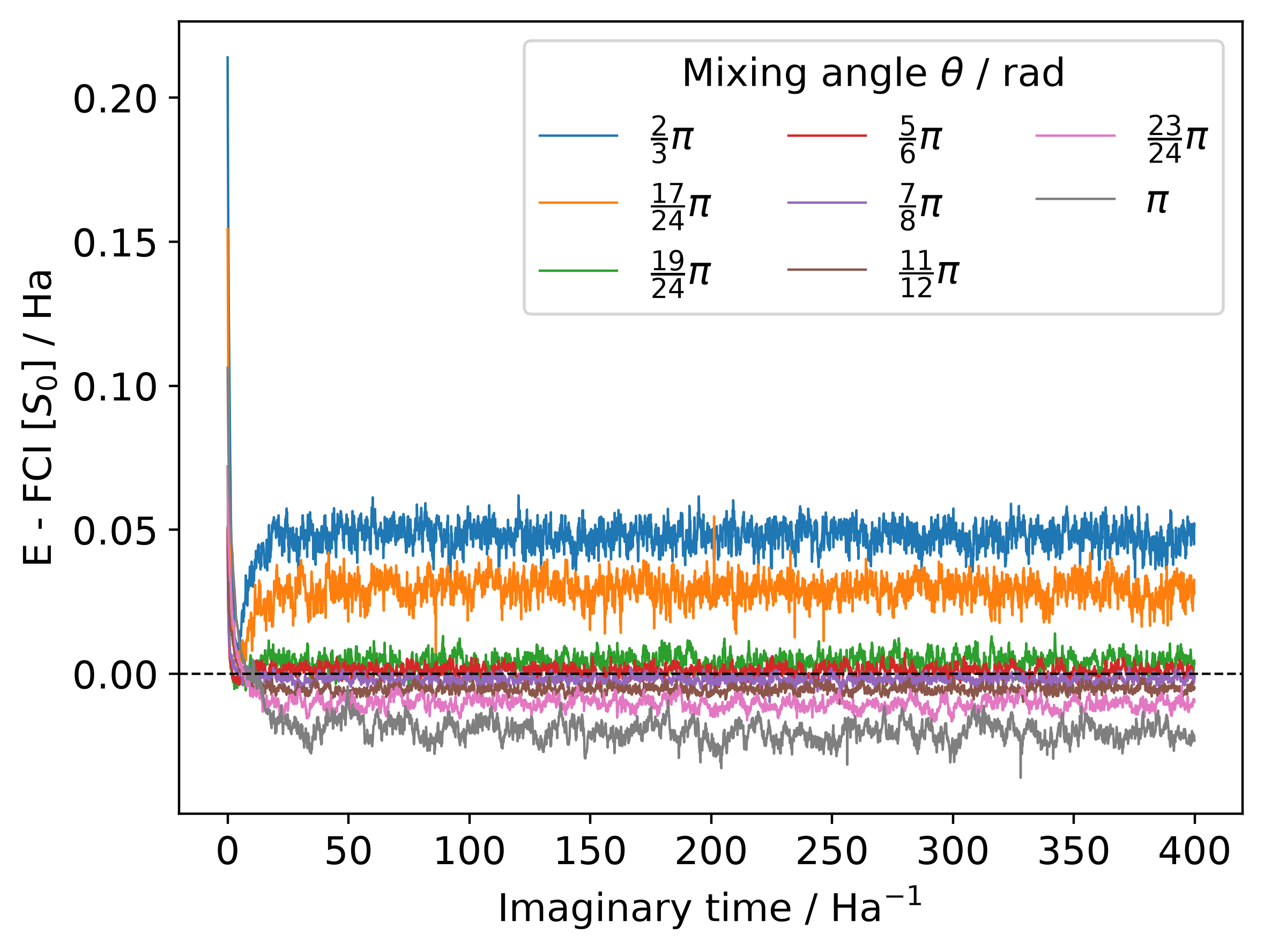}
    \caption{}
  \end{subfigure}
    \begin{subfigure}[b]{0.48\textwidth}
    \centering
    \includegraphics[width=\linewidth]{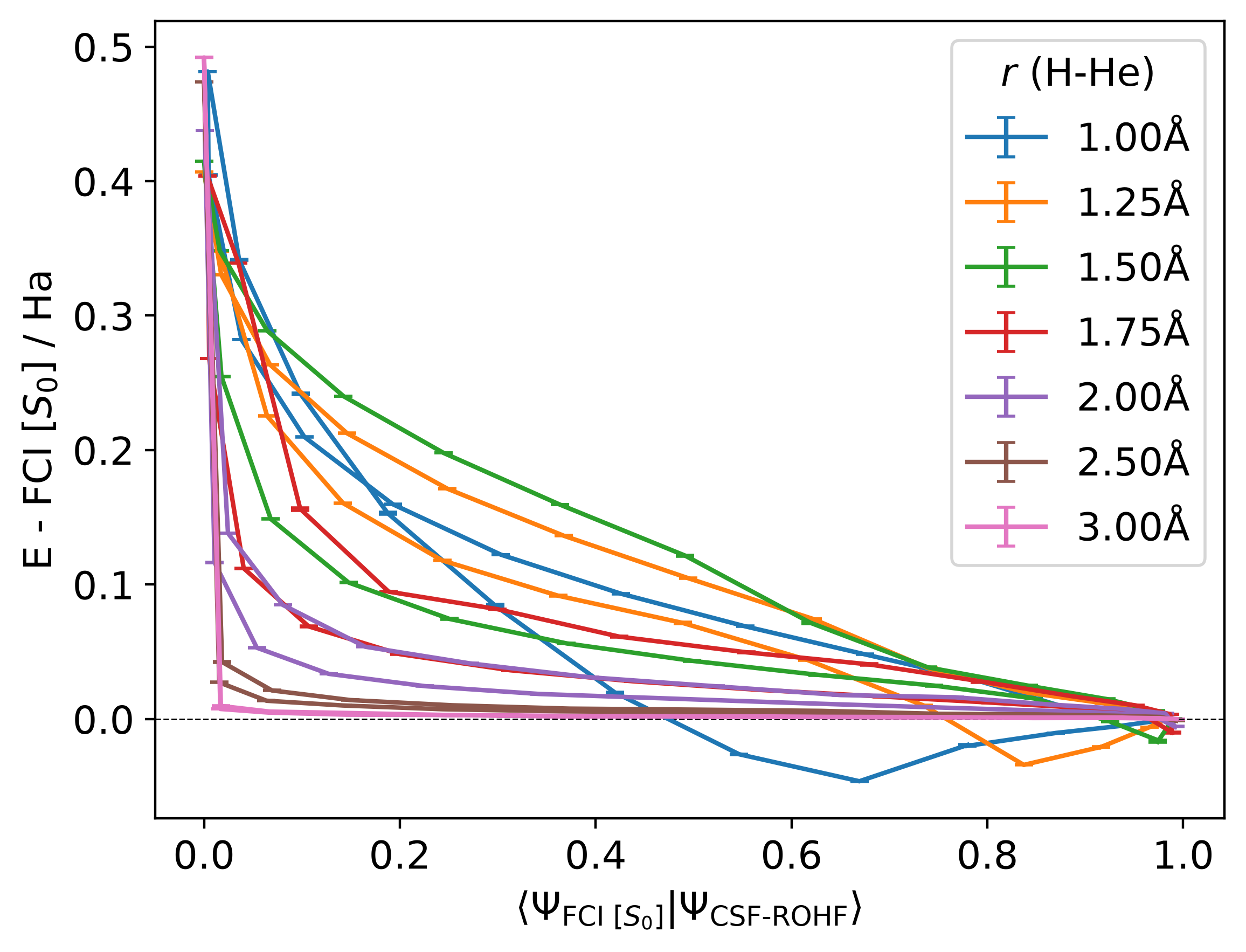}
    \caption{}
  \end{subfigure}
  \caption{\footnotesize (a) Energy differences between ph-AFQMC and FCI, and (b) overlaps between the trial wavefunction and FCI, as a function of $\theta$ of the trial at various \ce{H-He-H} bond lengths (note that intermediate points between 2\AA{} and 3\AA{} have been omitted for clarity).
  (c) Representative ph-AFQMC trajectories at eight different values of $\theta$ for $r(\ce{H-He})=\SI{1}{\angstrom}$ near the highest FCI overlap identified in panel (b). (d) Energy differences between ph-AFQMC and FCI plotted against the trial–-FCI overlap as $\theta$ is varied, using the corresponding energy differences from panel (a) and overlaps from panel (b).}
  \label{fig:hheh:theta_data}
\end{figure}

The energy deviations from FCI of the ph-AFQMC energies as a function of trial mixing angle is shown in \autoref{fig:hheh:theta_data}(a). 
The peaks corresponding to maximal ph-AFQMC energy error occur at $\theta$ values consistent with a near-zero overlap of the trial with the FCI ground state (the overlaps are shown separately in \autoref{fig:hheh:theta_data}(b)).  The shift in the peak position as the \ce{H-He} distances are stretched is consistent with the trial--FCI overlap trend, with the idealized ratio of open- to closed-shell character shifting towards larger $\theta$ with increasing bond length. %The peaks in energy errors (where AFQMC maximally deviates from the FCI energy)  
A related trend is observed in the imaginary-time ph-AFQMC trajectories for the $\SI{1}{\angstrom}$ system shown in \autoref{fig:hheh:theta_data}(c). For more CSF-ROHF-dominated trials (i.e., those with more open-shell character) we once again observe a shoulder followed by an artificially low energy plateau below the FCI energy, whereas trial states with too much closed-shell character initially decay normally but then sharply rise before plateauing at an artificially high value relative to FCI. These observations suggest that trials with too much open-shell character will systematically lead to ph-AFQMC energies below the exact value, potentially due to the sampling bias mechanism discussed above, whereas trials with too much closed-shell character lead to ph-AFQMC energies that are too positive, likely because the more closed-shell trials leads to the (full or partial) sampling of an excited eigenstate.  \autoref{fig:hheh:theta_data}(d) collects all of the data in panels (a) and (b), and shows that, in general, the ph-AFQMC energy accuracy is correlated with the overlap between the trial wavefunction and exact ground-state singlet, though the precise dependence is bond distance-dependent.
Notably, the ph-AFQMC energy accuracy becomes much less sensitive to the trial--FCI overlap for larger bond lengths, as demonstrated by the near step-like behavior for the energy error at $r(\ce{H-He}) = \SI{3.0}{\angstrom}$ in \autoref{fig:hheh:theta_data}(d).

\subsection{TS12 set}
\subsubsection{Assessment of CSF-ROHF trials in the  complete basis set limit}
The TS12 set of singlet-triplet gaps in $p$-block atoms, diatomics, and hydrides has previously been studied with a range of methods, including ph-AFQMC with various trials\cite{BaoGagliardiTruhlar,sheeSingletTripletEnergy2019,leeTwoSinglereferenceApproaches2019, MazziottiDiradicaloids,aiShortrangeScreenedDensity2023}. Lee et al.\  identified the relative success of ph-AFQMC using complex Restricted Hartree--Fock (cRHF) trials in comparison to RHF and Unrestricted Hartree--Fock (UHF) for the singlet states with multi-reference character.\cite{leeUtilizingEssentialSymmetry2020} %, explained as a manifestation of 'essential' symmetry breaking. 
We build directly on their work by initially considering a CSF-ROHF trial for the singlet and re-using their published UHF trial ph-AFQMC triplet values to obtain a singlet--triplet gap extrapolated to the complete basis set (CBS) limit that can be directly  compared with  experimental values. We have used the published geometries and experimental values from Ref. \citenum{leeTwoSinglereferenceApproaches2019} and the aug-cc-pVTZ and aug-cc-pVQZ\cite{Dunning1989, kendall1992a, woon1993a} basis sets throughout this section.

\begin{figure}
  \centering
  \includegraphics[width=0.6\linewidth]{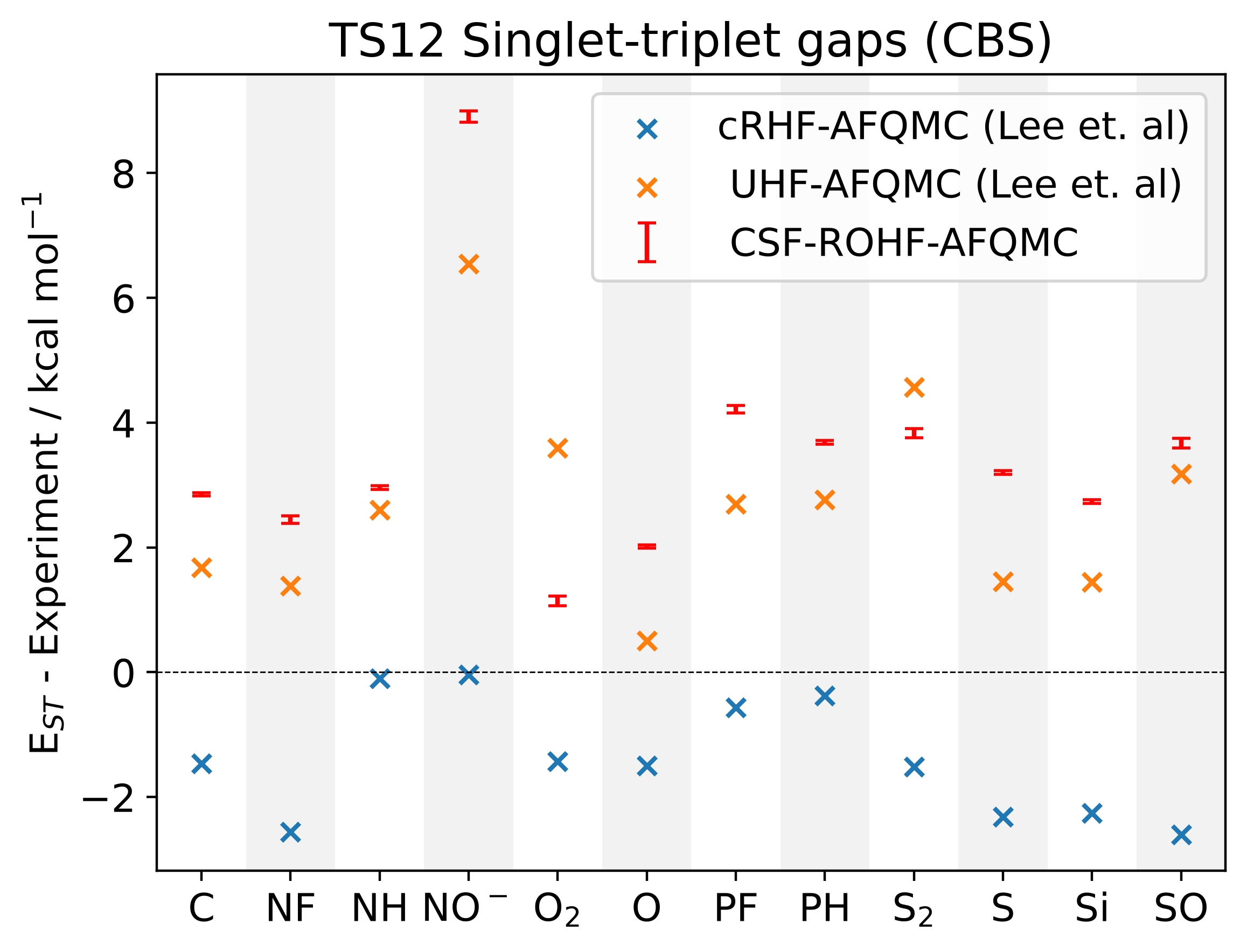}
  \caption{\footnotesize CBS-extrapolated (aug-cc-pVTZ/aug-cc-pVQZ) singlet-triplet gaps in kcal/mol, relative to experiment for the TS12 set. The UHF trial ph-AFQMC energies from Ref. \citenum{leeUtilizingEssentialSymmetry2020} were used throughout for the triplet states.}
  \label{fig:ts12:cbs}
\end{figure}

All of the CSF-ROHF wavefunctions have a 2-determinant form (with a spin-coupling pattern of $+-$) similar to the one used for \ce{H-He-H} above. As shown in \autoref{fig:ts12:cbs}, use of CSF-ROHF trials for the singlet states leads to systematically overestimated ph-AFQMC singlet-triplet gaps. 
This deviation arises because the singlet energy is artificially too high, as the same UHF trial ph-AFQMC triplet energy from Ref.~\citenum{leeUtilizingEssentialSymmetry2020} is used to compute singlet-triplet gaps for both cRHF-AFQMC and CSF-ROHF-AFQMC.
Representative statistical quantities for the CSF-ROHF-AFQMC singlet-triplet gap relative to experiment across this dataset are a Root Mean Square Deviation (RMSD) of 3.92 kcal/mol and a Mean Signed Deviation (MSD) (equivalent to the Mean Absolute Deviation (MAD) as all gaps are too large) of 3.47 kcal/mol. In comparison, use of cRHF trials for the singlets gives an RMSD of 1.71, MAD of 1.41, and MSD of $-1.41$ kcal/mol, and UHF trials give an RMSD of 3.11, MAD of 2.71, and MSD of 2.71 kcal/mol.  The CSF-ROHF trial for the singlet states evidently leads to worse accuracy in the singlet-triplet gaps compared to both cRHF and UHF trials.
Two probable causes of this relatively poor level of accuracy are $L_z$ symmetry breaking in the trial state and/or suboptimal orbitals. Both are explored in detail below.

\subsubsection{$L_z$ symmetry}
\begin{figure}
  \centering
  \includegraphics[width=0.6\linewidth]{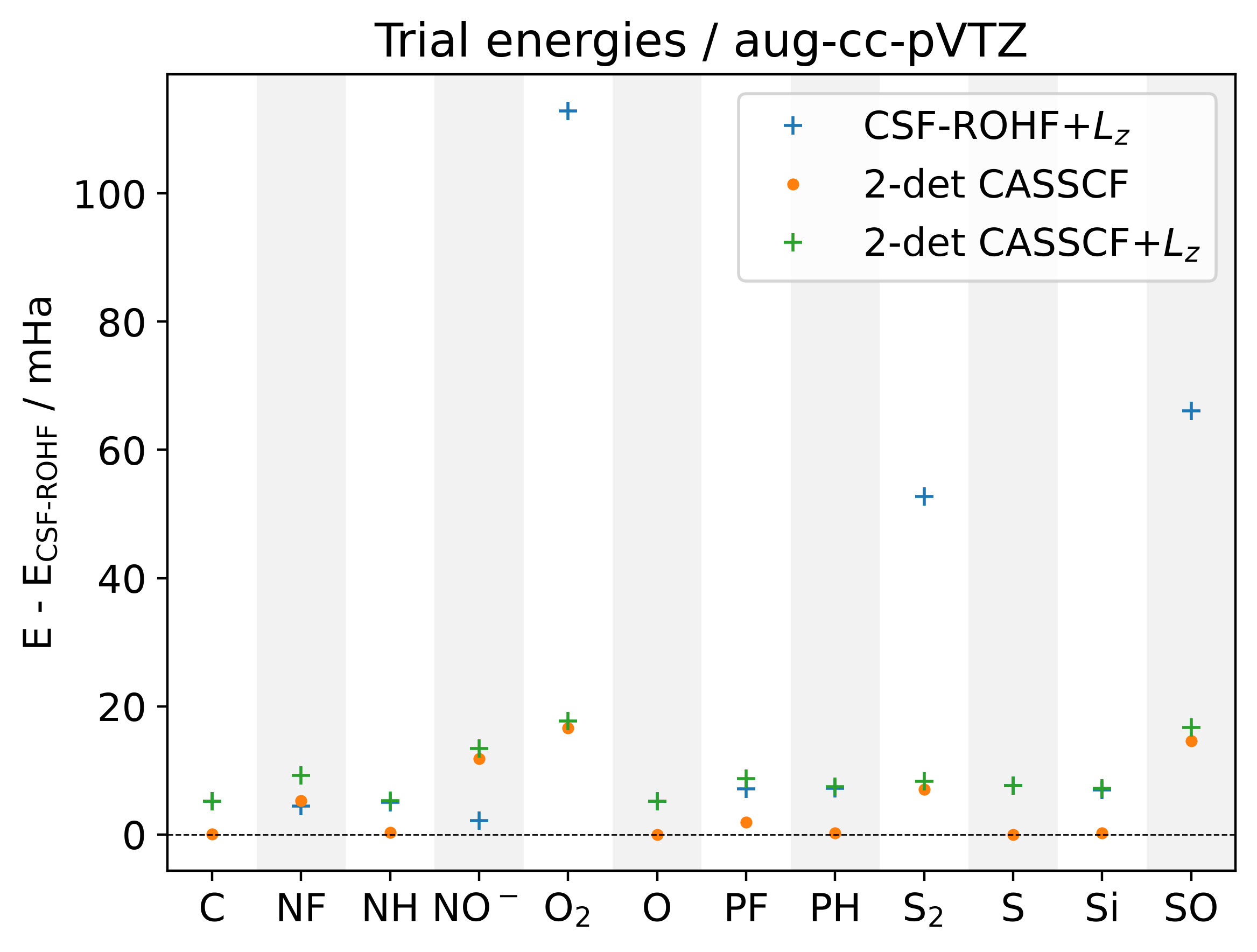}
  \caption{\footnotesize Energy shifts (in mHa) relative to the CSF-ROHF energies associated with $L_z$-purification of CSF-ROHF states, truncated 2-determinant CASSCF wavefunctions and their associated $L_z$-purified versions in the aug-cc-pVTZ basis.}
  \label{fig:ts12:csf_lz_gap}
\end{figure}

One possible origin of these large errors is that, unlike cRHF wavefunctions, CSF-ROHF states using real orbitals can break orbital angular momentum symmetry.  As discussed in detail in \autoref{appendix:lz}, the target (lowest-energy) singlet state for, e.g., the \ce{O2} molecule has term symbol $^1\Delta_\text{g}$, and must contain both closed-shell and open-shell contributions (when using a basis of real orbitals) to be an eigenfunction of the $\hat{L}_z$ operator with eigenvalue $m_l = 2$.  
%\begin{equation}
%    ^1\Delta_\text{g} = \frac{1}{2}\Big((\abs{x\bar{x}} - \abs{y\bar{y}}) \pm \mathrm{i} (\abs{x\bar{y}} + \abs{y\bar{x}})\Big)
%\end{equation}  %In this context, symmetry breaking refers to a failure to enforce an observable operator's (i.e. one which commutes with the Hamiltonian) value by wavefunction construction. A helpful analogy is the $\hat{S}^2$ spin-symmetry breaking observed in UHF in comparison to RHF wherein a choice 
We can enforce $L_z$ symmetry on an existing singlet CSF-ROHF solution by introducing two closed shell configurations and allowing for complex-valued CI coefficients:
\begin{equation}
   \ket{\text{CSF-ROHF+$L_z$}} =\frac{1}{\sqrt2} \left[\frac{1}{\sqrt{2}}(\ket{\text{core}\cdots 20} - \ket{\text{core}\cdots 02}) + i\underbrace{\frac{1}{\sqrt{2}}(\ket{\text{core}\cdots \uparrow \downarrow}  +\ket{\text{core}\cdots \downarrow \uparrow})}_\text{CSF-ROHF}\right]
   \label{eq:CSFts12}
\end{equation}
We refer to this wavefunction structure as CSF-ROHF+$L_z$.
Note that the $L_z$-purified wavefunction above is higher in energy than the original CSF-ROHF state (a clear manifestation of L\"owdin's symmetry dilemma\cite{paldusHartreeFockStability1985, Lykos1963} where enforcing a symmetry on an approximate wavefunction carries an energetic penalty). This energetic penalty varies significantly in value but is most acute in O$_2$, S$_2$, and SO where $L_z$ restoration carries an energetic cost of 113, 53, and 66 mHa, respectively (\autoref{fig:ts12:csf_lz_gap}). We remark that group 16 diatomics specifically appear to be the most sensitive and, even though \ce{NF} and \ce{NO^-} are isoelectronic to \ce{O2}, the energetic effect of the $L_z$ purification is an order of magnitude smaller in these cases.

\begin{figure}[htb]
  \centering
  \includegraphics[width=0.99\linewidth]{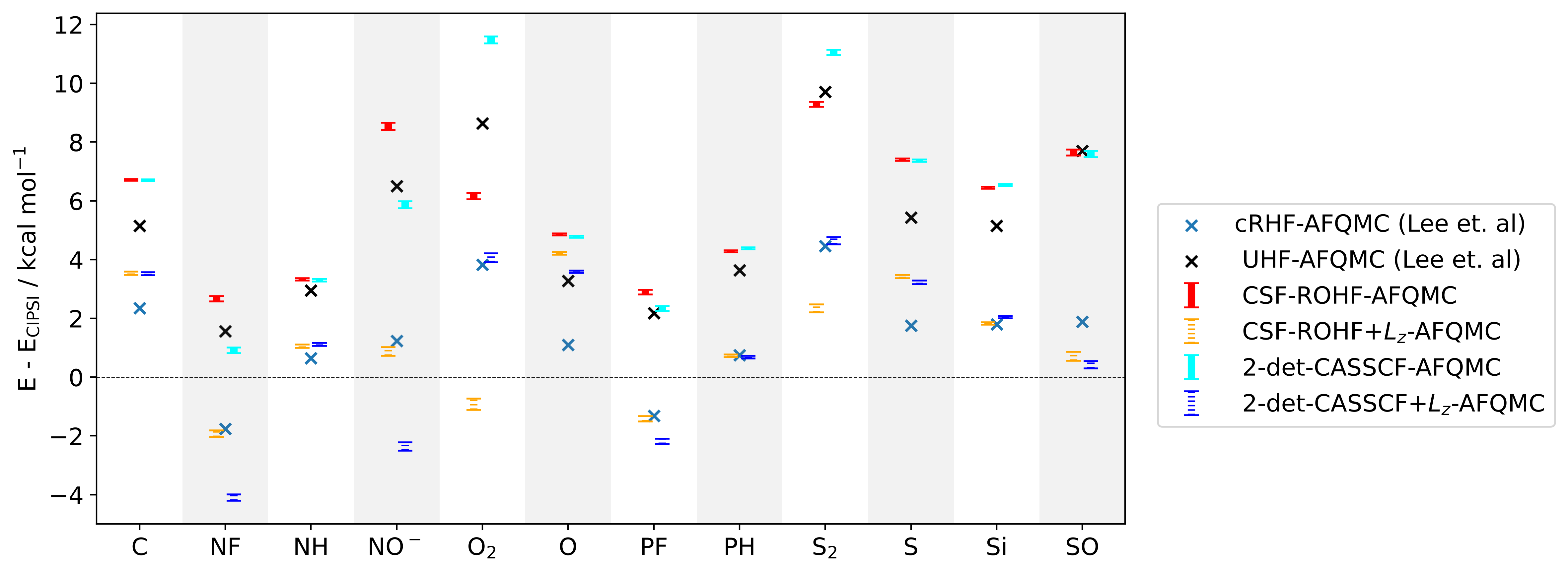}
  \caption{\footnotesize Singlet energies relative to the extrapolated CIPSI reference for the TS12 set with ph-AFQMC using both CSF-like (either CSF-ROHF or 2-determinant truncated CASSCF) and orbital angular momentum-purified (+$L_z$) trials in the aug-cc-pVTZ basis.}
  \label{fig:ts12:cas-lz}
\end{figure}

To assess the use of these $L_z$-adapted CSF-ROHF wavefunctions as ph-AFQMC trial states, we perform calculations in the aug-cc-pVTZ basis, %(instead of extrapolating to the CBS limit), 
and use reference values obtained from CIPSI calculations instead of experimental values, allowing us to verify the accuracy of the singlet state energies directly. %This is a deliberate choice to highlight the trends in the raw singlet energies to eliminate any possible error in the triplet values as we alter the trial.
The data in \autoref{fig:ts12:cas-lz} show that the ph-AFQMC singlet energies using CSF-ROHF$+L_z$ trials are consistently lower than those using CSF-ROHF trials. This is reflected in the statistical metrics compared to CIPSI reference  energies across the set, shown in \autoref{table:ts12:cipsi_errors}.  Without $L_z$ projection the CSF-ROHF trial ph-AFQMC RMSD, MAD, and MSD are 6.23, 5.85, and 5.85 kcal/mol, respectively; the CSF-ROHF+$L_z$ trial significantly improves these metrics of the corresponding ph-AFQMC calculations to 2.24, 1.91, and 1.20 kcal/mol.  The latter ph-AFQMC protocol with $L_z$ purified CSF-ROHF trials is comparable in accuracy to ph-AFQMC with cRHF trials, which is not surprising given that the cRHF wavefunction can be expanded exactly in the 4-determinant basis of  Eq.~\eqref{eq:complexCSF} with the same functional form as Eq.~\eqref{eq:CSFts12} (shown in \autoref{appendix:lz}).

\begin{table}[htb]
  \caption{\footnotesize Mean absolute deviation (MAD), mean signed deviation (MSD) and root mean square deviation (RMSD), from the extrapolated CIPSI energies for the open-shell singlet ground states in aug-cc-pVTZ basis for the TS12 set, in kcal/mol. For our ph-AFQMC calculations, the largest ph-AFQMC statistical error was 0.19 kcal/mol, with an average error of 0.08 kcal/mol.}
  \label{table:ts12:cipsi_errors}
  \begin{tabular}{r  c c c}
  \hline\hline
    Method & RMSD & MAD & MSD \\
    \hline
    cRHF-AFQMC (Lee et al.)\cite{leeUtilizingEssentialSymmetry2020}  & 2.21 & 1.91 & 1.39 \\
    UHF-AFQMC (Lee et al.)\cite{leeUtilizingEssentialSymmetry2020}   &  5.72 & 5.16 & 5.16\\
    \hline
    CSF-ROHF-AFQMC           & 6.23 & 5.85 & 5.85 \\
    CSF-ROHF+$L_z$-AFQMC     & 2.24 & 1.91 & 1.20 \\
    2-det-CASSCF-AFQMC       & 6.75 & 6.02 & 6.02 \\
    2-det-CASSCF+$L_z$-AFQMC & 2.98 & 2.66 & 1.22 \\
      \hline\hline
  \end{tabular}
\end{table}

A few system-specific comments are in order.  
In NF, NO$^-$, PF, PH, and Si the CSF-ROHF+$L_z$ trial ph-AFQMC energies are very close to those of cRHF trial ph-AFQMC. 
In the three systems exhibiting relatively large $L_z$ energy shifts (\ce{O2}, \ce{S2}, and \ce{SO} as noted above) the CSF-ROHF+$L_z$ trial ph-AFQMC performs better than when the cRHF trial is used; yet in \ce{C}, \ce{NH}, \ce{O}, and \ce{S} the $L_z$ purified CSF-ROHF leads the ph-AFQMC to higher (and less accurate) energies vs the cRHF trial calculations. This mixed picture suggests that ph-AFQMC and the choice of trial wavefunction have a high degree of sensitivity to the underlying orbital set, which we shall explore further in the next section.

\subsubsection{Orbital sensitivity of CSF-ROHF-like trials in ph-AFQMC}

%We now consider the effect of changing the orbitals.
While CSF-ROHF minimizes the energy with respect to the orbitals used in the CSF, another reasonable set of candidate orbitals would be those obtained from a state-specific CASSCF calculation. Note that all of the CASSCF solutions we have identified in this section (see SI for active space information) were dominated by the CSF configuration. %(consistent with the targeting of the same open-shell singlet state). 
To make the comparison more direct, i.e., to isolate the effect of the orbitals only, we have intentionally discarded most of the CASSCF wavefunction CI expansion, truncating it to keep only the 2 determinants of character corresponding directly to the CSF-ROHF state. %This wavefunction could equally be considered an unoptimised CSF-ROHF.

Across the TS12 set using the aug-cc-pVTZ basis, we observe that the 2-determinant CASSCF trials are all equal or higher in energy than the CSF-ROHF states (\autoref{fig:ts12:csf_lz_gap}); this result is unsurprising as the CASSCF orbitals were optimised in the field of multiple configurations that have been discarded,
and the CSF-ROHF solutions are the variational minima for these spin-coupled configurations.  %Critically, this does support the idea that we have located valid low-lying CSF-ROHF solutions and not some artificial local minima. 
For the rest of this discussion we will focus on the systems where the CSF-ROHF and 2-determinant CASSCF energies disagree ---  specifically \ce{NF}, \ce{NO-}, \ce{O2}, \ce{PF}, \ce{S2}, and \ce{SO}.
Using these CASSCF-orbital CSF-ROHF-like trials in ph-AFQMC calculations (denoted 2-det-CASSCF-AFQMC), we once again observe a systematic overestimation of the singlet energy (\autoref{fig:ts12:cas-lz}). The set-wide statistics in \autoref{table:ts12:cipsi_errors} suggest a marginally worse overall performance than the CSF-ROHF trial states. 
For NF, NO$^-$, and PF the ph-AFQMC energies are lower with CASSCF-optimized orbitals (in this context, more accurate relative to the CIPSI reference), providing examples of when a higher energy trial has led to a lower ph-AFQMC energy. However, this trend is not universal. In \ce{O2} and \ce{S2} the energies are higher (i.e., worse as they have a higher deviation from CIPSI) and in SO the energy is almost the same (within error bars).

$L_z$ symmetry can also be restored in these truncated CASSCF wavefunctions in a similar way as explained above for the CSF-ROHF wavefunctions; we refer to this trial protocol as 2-det-CASSCF+$L_z$. 
The full data set is shown in \autoref{fig:ts12:cas-lz}. Overall, given the statistical results in \autoref{table:ts12:cipsi_errors}, it would appear that the CSF-ROHF optimized orbitals lead to better ph-AFQMC energies than their CASSCF-optimized counterparts, especially when performing $L_z$ symmetry restoration. This is an encouraging finding, given the CSF-ROHF-based protocol's advantage of significantly lower computational cost especially if large active spaces are required.

\subsection{Twisted Ethene}
We next consider two non-equilibrium conformers of ethene (also known as ethylene), C$_2$H$_4$, obtained by twisting around the $\phi_\text{H-C-C-H}$ dihedral angle. For the other geometric parameters we have used $r_\text{C-C} = \SI{1.33381}{\angstrom}$, $r_\text{C-H} = \SI{1.08035}{\angstrom}$, and $\theta_\text{H-C-H}=117.23174^\circ$ based on the QUEST Dataset geometry.\cite{verilQUESTDBDatabaseHighly2021} Ethene has a rich photochemistry that has been the subject of many previous computational studies.\cite{buenkerPotentialEnergyDipole1980, petrongoloNonadiabaticTreatmentIntensity1982, serrano-andresAccurateMolecularOrbital1993, barbattiPhotochemistryEthyleneMultireference2004, wangTwistedTripletEthylene2014,saadeExcitedStateSpecificCASSCF2024}
At $\phi_\text{H-C-C-H} = 90^\circ$ the system is a perfect diradical with unpaired electrons on either carbon in the orthogonal $p$ orbitals, 
leading to a degenerate pair of RHF solutions.\cite{Burton2018}
Unlike the systems in the TS12 set discussed above, this system does not have any continuous $L_z$ symmetry that needs to be conserved, suggesting that a single CSF trial may be more appropriate for the open-shell singlet.

\begin{table}[b!]
  \caption{\footnotesize Singlet-triplet ($E_\text{S} - E_\text{T}$) gaps for two conformers of twisted ethene using various methods, in kcal/mol. Note that the CSF-ROHF and CSF-ROHF trial ph-AFQMC methods use the corresponding UHF and UHF trial ph-AFQMC energies for the triplet state, respectively. 
  }
  \label{table:ethene2}
  \begin{tabular}{r  c  c}
      \hline\hline
    Method & $\phi_\text{H-C-C-H}  = 82.5 ^\circ$ & $\phi_\text{H-C-C-H}  = 90.0 ^\circ$ \\
    \hline
    UHF         & $-4.22$ & $-2.96$ \\
    CSF-ROHF         & $+5.88$ & $+5.83$ \\
    \hline
    RCCSD       & $+5.57$ & $+17.61$ \\
    RCCSD(T)    & $-1.67$ & $+5.42$ \\
    UCCSD   & $-2.77$   & $-0.63$\\
    UCCSD(T) & $-2.78$  & $-0.47$ \\
    \hline
    UHF trial ph-AFQMC   & $-5.26\pm0.04$ & $+0.04\pm0.03$\\
    CSF-ROHF trial ph-AFQMC   & $-5.72\pm0.04$ & $+0.10\pm0.03$ \\
    \hline
    CIPSI Reference & $-4.92\pm 0.07$ & $-0.87\pm0.10$\\
    \hline\hline
  \end{tabular}
\end{table}

\autoref{table:ethene2} shows the singlet-triplet gaps ($E_\text{S} - E_\text{T}$) for dihedral angles of 82.5$^\circ$ and 90.0$^\circ$ computed with various methods, along with reference CIPSI values.  The data presented were obtained in the aug-cc-pVDZ\cite{Dunning1989, kendall1992a} basis, motivated by the relatively high cost of the CIPSI  calculations. CSF-ROHF predicts gaps of incorrect sign in both cases, as does RCCSD.  At the 82.5$^\circ$ twist angle, RCCSD(T) is of the correct sign but is still over 3 kcal/mol from the reference value; while UCCSD and UCCSD(T) are of the correct sign, both are roughly 2 kcal/mol away from the reference value.  The ph-AFQMC methods, using either UHF or CSF-ROHF as trials, are closest to the CIPSI gap among all of the correlated methods investigated, leading to sub kcal/mol accuracy.

Regarding the ethene conformation with 90.0$^\circ$ dihedral twist, the accuracy of RCCSD and RCCSD(T) becomes substantially worse (relative to the 82.5$^\circ$ twist angle), with the latter more than 6 kcal/mol away from the CIPSI reference gap. 
At this geometry, two degenerate RHF solutions cross,\cite{Burton2018} creating an unphysical cusp in the RHF and corresponding RCCSD energies.\cite{Gulania2021,Musial2011}
The use of unrestricted HF as reference provides a significant improvement to both CCSD and CCSD(T), with both methods achieving sub-kcal/mol accuracy.  Both ph-AFQMC methods are roughly within statistical error of one another and once again achieve sub-kcal/mol accuracy compared to the reference singlet--triplet gap. 
Selected total energies are shown in \autoref{table:ethene1}, highlighting that essentially exact values are obtained from ph-AFQMC with both UHF and CSF-ROHF trial wavefunctions.

\begin{table}[htb]
  \caption{\footnotesize Total energies (Ha) for the two conformers of twisted ethene using various methods. All CIPSI extrapolation uncertainties and ph-AFQMC statistical error bars are less than 1 mHa. For the triplet state, UHF and ROHF are equivalent; also, a triplet CSF-ROHF calculation would have only one determinant and be equivalent to ROHF in this system.
  }
  \label{table:ethene1}
  \begin{tabular}{r  c c  c c}
      \hline\hline
       & \multicolumn{2}{c}{\underline{$\phi_\text{H-C-C-H}  = 82.5 ^\circ$}} & \multicolumn{2}{c}{\underline{$\phi_\text{H-C-C-H}  = 90 ^\circ$}} \\
    Method & Singlet & Triplet & Singlet & Triplet\\

    \hline
%    RHF           & $-77.897 5$ & --- & $-77.871 0$ & --- \\
    RCCSD         & $-78.237 1$ & --- & $-78.218 6$ & --- \\
    RCCSD(T)      & $-78.256 3$ & --- & $-78.245 6$ & --- \\
%    \hline
%    UHF           & $-77.966 3$ & $-77.959 6$ & $-77.964 9$ & $-77.960 2$ \\
    UCCSD         & $-78.250 4$ & $-78.246 0$ & $-78.247 6$ & $-78.246 6$ \\
    UCCSD(T)      & $-78.258 0$ & $-78.253 6$ & $-78.255 0$ & $-78.254 3$ \\
    \hline
    % UHF trial ph-AFQMC           & $-78.264 207	\pm0.000 041$ & $-78.255 830 \pm 0.000 038$ & $-78.256 341	\pm0.000 033$ & $-78.256 412\pm 0.000 036$ \\
    UHF trial ph-AFQMC           & $-78.264 2	$ & $-78.255 8$ & $-78.256 3	$ & $-78.256 4$ \\
%    CSF-ROHF      & $-77.950 2$ & --- &  $-77.950 9$ & ---\\
    % CSF-ROHF trial ph-AFQMC      & $-78.264 938	\pm 0.000 043$ & --- & $-78.256 255	\pm 0.000 039$ & --- \\
    CSF-ROHF trial ph-AFQMC      & $-78.264 9	$ & --- & $-78.256 3$ & --- \\
    \hline
    % CIPSI         & $-78.262 679\pm 0.000 105$ & $-78.254836\pm0.000 013$ & $-78.256 898 \pm 0.000 154$ & $-78.25550912\pm	0.000 038$ \\
    CIPSI         & $-78.262 7$ & $-78.254 8$ & $-78.256 9 $ & $-78.255 5$ \\

    \hline\hline
  \end{tabular}
\end{table}

%\newpage
\subsection{Bimetallic iron-sulfur complexes}
Lastly, we turn to the synthetic [Fe$_2$S$_2$(SCH$_3$)$_4$]$^{
2-}$ complex, which we will henceforth refer to as [2Fe--2S]$^{2+}$.  We will present data for the 30e20o active space model from Chan and coworkers,\cite{liSpinProjectedMatrixProduct2017} alongside a full space (within the usual frozen-core approximation) treatment in the cc-pVDZ basis.  Both used the DFT-optimized geometry from Ref. \citenum{sharmaLowenergySpectrumIron2014}.
In this case, the singlet CSF-ROHF ground state corresponds to the spin-coupling pattern $+++++-----$, representing two high-spin $S=\tfrac{5}{2}$ states on each \ce{Fe} centre antiferromagnetically coupled  to an overall singlet.
We also consider CASSCF calculations with varying active spaces to assess the importance of beyond-CSF configurations in the quality of the trial state.

\begin{table}[htb]
  \caption{\footnotesize Singlet energies of the Li \& Chan \cite{liSpinProjectedMatrixProduct2017} [2Fe--2S]$^{2+}$ \ddv{30e20o} active space model with various methods. The 10e10o CASSCF space consists of the ten Fe 3$d$ orbitals, while the 22e16o space adds the six bridging sulfur $p$ orbitals.}
  \label{table:fes-activespace}
  \begin{tabular}{l  l }
  \hline\hline
    Method & Singlet energy / Ha\\
    \hline
    CSF-ROHF & $-116.486 6$\\
    CASSCF(10e,10o) & $-116.5172$\\
    CASSCF(22e,16o) & $-116.595 7$\\
    \hline
    CSF-ROHF trial ph-AFQMC & $-116.6205 \pm 0.000 3$\\
    CASSCF(10e,10o) trial ph-AFQMC & $-116.5896 \pm 0.0002$\\
    \hline
    DMRG Reference Li \& Chan (2017)\cite{liSpinProjectedMatrixProduct2017} & $-116.6056$\\
    \hline\hline
  \end{tabular}
\end{table}

Total energies of the singlet ground state in the active space model are shown in \autoref{table:fes-activespace}, with the DMRG energy from Ref. \citenum{liSpinProjectedMatrixProduct2017} used as a reference.  As expected, we find a systematic energy lowering from CSF-ROHF to CASSCF(10e10o) (with the active space initialized from the 10 localized Fe 3$d$ orbitals) to CASSCF(22e16o).  However, the energy from ph-AFQMC with the CSF-ROHF trial is overly negative by $\sim$15 mHa, reminiscent of the results for the \ce{H-He-H} system at shorter bond lengths. This is chemically consistent, given that in both cases (short-bond length HHeH and [2Fe--2S]$^{2+}$) the exact eigenstate is expected to exhibit substantial contributions from `ionic' configurations, which are missing in the CSF-ROHF trials.  Note that our previous study of this active space considered various UHF trial wavefunctions, the best of which yielded a ph-AFQMC energy error of 13.9 ± 0.2 mHa \textit{above} the DMRG reference.\cite{danilovSelectingOptimalUnrestricted2026} 
Using a CASSCF(10e10o) trial wavefunction, ph-AFQMC undercorrelates to give an energy that is higher than that of DMRG by roughly 16 mHa; in this ph-AFQMC calculation we kept 99.95\% of the CASSCF wavefunction as the trial and utilized the CASSCF natural orbital basis.  These results are qualitatively consistent with another study\cite{huangGPUacceleratedAuxiliaryfieldQuantum2024} on the same active space model using ph-AFQMC with heat-bath selected CI trial wavefunctions, which demonstrated convergence to the exact energy with a million determinants \emph{from above} when natural orbitals (from a near-exact DMRG calculation) were used, and overly-negative energies compared to the exact energy when localized orbitals were used (which we presume to be the high-spin localized DFT orbitals used to construct the active space).

The $\mathcal{O}(N^4)$ scaling of the CSF-ROHF procedure enabled us to also consider the full-space [2Fe--2S]$^{2+}$ molecule in the cc-pVDZ basis.\cite{balabanov2005a, balabanov2006a, kendall1992a, woon1993a, Dunning1989}  The energy gap between the %undecatet (11-fold multiplicity) 
$S=5$ `high-spin' state and the low-spin singlet state is shown in \autoref{table:fes-fullspace}, using various methods.  Treating the former with ROHF and using CSF-ROHF for the low-spin state led to a gap of $-2.6$ kcal/mol, which is of opposite sign to the reference values taken from previous calculations including stochastic CASSCF(22e26o),\cite{Dobrautz2021} DMRG on the 30e20o active space,\cite{sharmaLowenergySpectrumIron2014} and full-space broken-symmetry CCSD(T) based on the $J$ coupling constant computed in the comparable cc-pVDZ basis.\cite{schurkusTheoreticalPredictionMagnetic2020}  ph-AFQMC with these ROHF and CSF-ROHF trials successfully restores the expected sign of the spin-gap, but the value obtained ($\sim$51.8(3) kcal/mol) seems much too large.  Given that we compute the gap as high-spin minus low-spin, the energy of the singlet state computed from ph-AFQMC with the CSF-ROHF trial is likely to be overly negative, consistent with our findings in the active space model above.  Indeed, using a CASSCF(10e10o) trial instead of the CSF-ROHF trial for the low-spin singlet state lowered the computed spin-gap to 29.5(4) kcal/mol, which agrees more closely with the aforementioned previous results.  A more comprehensive investigation of the total energies of both the $S=0$ and $S=5$ states  with various trial wavefunctions in this basis will be the topic of future work. 
\begin{table}
  \caption{\footnotesize $E(S=5) - E(S=0)$ gaps for full space [2Fe--2S]$^{2+}$ calculations. The $\sim$ indicates that the value was estimated from plots, as the raw numerical energies were not available.}
  \label{table:fes-fullspace}
  \begin{tabular}{l  l   c}
    \hline\hline
  High-spin ($S=5$) method & Low-spin ($S=0$) method & Gap (kcal/mol)\\
  \hline
  ROHF & CSF-ROHF & -2.603\\
  ROHF & CASSCF(10e10o) & 5.180\\
  \hline
  ROHF trial ph-AFQMC & CSF-ROHF trial ph-AFQMC & $51.815 \pm 0.286$\\
  ROHF trial ph-AFQMC & CASSCF(10e10o) trial ph-AFQMC & $29.508 \pm 0.407$ \\
  \hline
  \multicolumn{2}{l }{Dobrautz et al. \cite{Dobrautz2021} FCIQMC-CASSCF(22e26o)} & $\sim$24\\
  \multicolumn{2}{l }{Sharma et al. \cite{sharmaLowenergySpectrumIron2014} 30e20o Active space DMRG} & $\sim$22\\
  \multicolumn{2}{l }{Schurkus et al. \cite{schurkusTheoreticalPredictionMagnetic2020} BS-CCSD(T) (idealised gap based on J coupling)} & 16\\
  \hline\hline
  \end{tabular}
\end{table}

\section{Discussion and conclusions}

Strongly correlated open-shell states, as encountered along bond-dissociation coordinates, in di-/poly-radicaloid molecules and many polynuclear transition metal clusters, are challenging to describe accurately and in a scalable manner using approximate electron correlation models. 
In this work, we explored the use of single orbital-optimized Configuration State Functions (CSFs) as trial wavefunctions in   phaseless auxiliary-field quantum Monte Carlo (ph-AFQMC), and considered the symmetric stretch of \ce{H-He-H}, a set of small molecules with triplet ground states and the TS12 diradicaloid singlet state set, twisted ethene/ethylene which exhibits substantial diradical character, and a synthetic  di-ferric [2Fe--2S]$^{2+}$ complex in both a valence active space model and the full cc-pVDZ 
orbital space  (with a frozen-core approximation).    

CSF-ROHF wavefunctions are spin-pure (with regard to both $\hat{S}^2$ and $\hat{S}_z$) by construction, and can robustly capture strong correlations arising from spin entanglement, which, in the space of Slater determinants, is often the dominant source of multireference character.  Another advantageous feature of CSF-ROHF is the $\mathcal{O}(N^4)$ scaling with system size of orbital optimization, which does not exceed the $\mathcal{O}(N^4)$ per-sample scaling of canonical AFQMC in Gaussian basis sets.  We note that the number of determinants within a single CSF scales combinatorially with the number of unpaired electrons; this cost will carry over to the AFQMC as well. 
Yet for most systems of chemical and biological relevance, the number of spin-entangled orbitals is much smaller than the size of the full orbital space, preserving polynomial system size scaling in practice. As a relevant example, the polynuclear FeMo cofactor that catalyses nitrogen fixation contains eight metallic centers,\cite{danceElectronicDimensionsFeMoco2011} while most other biological transition metal clusters contain fewer metal centers.  
Furthermore, the number of determinants within a CSF can be significantly less than the equivalent minimal active-space expansion, with as much as a square-root reduction for a singlet state.

We find that CSF-ROHF can be an effective trial wavefunction for ph-AFQMC studies of systems with strongly localized unpaired electrons and low point-group symmetry, such as ethene twisted about its dihedral angle by 82.5$^\circ$ and 90$^\circ$.  Indeed, ph-AFQMC with CSF-ROHF trials led to sub kcal/mol accuracy vs CIPSI references for the singlet-triplet gaps, while both restricted and unrestricted CCSD(T) showed deviations in excess of 2 kcal/mol at the 82.5$^\circ$ twist angle.  However, CSF-ROHF does not appear to be a universally optimal trial wavefunction, as our study of other systems revealed a number of unexpected findings.  In particular, results for the TS12 set suggest that spin symmetry -- while obviously very important -- is not the \emph{only} important symmetry that an optimal trial wavefunction must respect in the case of diradicaloid singlet states of atoms and diatomic molecules.  In these cases, we restored $L_z$ symmetry in the trial wavefunction by augmenting the two-determinant open-shell singlet CSF with two more closed-shell determinants (and allowing the four CI coefficients to be complex-valued).  Without further orbital optimization, ph-AFQMC calculations with such trials yield singlet energies with an accuracy rivaling ph-AFQMC with complex restricted Hartree--Fock trial wavefunctions.    

The second significant shortcoming of ph-AFQMC with CSF-ROHF trial wavefunctions is revealed in our studies of \ce{H-He-H} at relatively short bond lengths and the [2Fe--2S]$^{2+}$ compound.  In both systems, the true wavefunction is known to exhibit both covalent/neutral (i.e., with unpaired localized electrons only on the H or Fe atoms for \ce{H-He-H} or [2Fe--2S]$^{2+}$, respectively) and ionic (i.e., with one or more H or Fe orbitals doubly-occupied) contributions.  This arises from the He- or S-mediated superexchange mechanism, in which the bridging atom/s can participate in correlated many-electron delocalization or charge-transfer processes.  However, the CSF-ROHF trial wavefunctions do not capture any ionic character, and in both short distance \ce{H-He-H} and [2Fe--2S]$^{2+}$ we observe overly negative ph-AFQMC energies (versus reference ground-state energies from FCI or bond dimension-converged DMRG).  This non-variational character has been observed in previous work, including studies of iron-sulfur cluster active space models with unrestricted Hartree--Fock trials, and has been attributed to a sampling bias that is distinct from the phaseless bias, yet is nevertheless caused by a suboptimal trial wavefunction.\cite{danilovSelectingOptimalUnrestricted2026}  The present analysis of the imaginary-time walker dynamics in the \ce{H-He-H} ($r=1$\AA) system is consistent with this mechanism. 

Taken together, while there are some specific cases where CSF-ROHF trials do lead to excellent accuracy, there is a clear need for more sophisticated trial wavefunctions especially for challenging classes of systems like polynuclear iron-sulfur clusters.  
We have shown that the performance of ph-AFQMC for such systems depends on the choice of orbitals and moreover the balance of covalent/neutral and ionic contributions in the trial wavefunction, in addition to spin symmetry preservation.  These findings should motivate and inform future studies of tractable and optimal trial wavefunction forms in constrained projector quantum Monte Carlo approaches.

\section{Computational details}\label{appendix:CompDetails}
We used PySCF v2.8.0\cite{sunRecentDevelopmentsPySCF2020} to generate the one and two electron integrals, and to perform CASSCF, Coupled Cluster, and FCI calculations. 
Orbitals were plotted using MultiWFN v2026.7.15\cite{luMultiwfnMultifunctionalWavefunction2012, luComprehensiveElectronWavefunction2024}

The \textsc{Quantel} software library\cite{Quantel} was used to generate the CSF-ROHF states. 
For \ce{HHeH}, the TS12 test set, and ethene, initial guesses were obtained using random orbital coefficients to ensure that the global energy minimum is located.
For the di-iron complex in the cc-pVDZ basis, we employed the initialization protocol outlined in Ref.~\citenum{Burton2025}. 
First, orbitals were obtained from a high-spin ROHF calculation with the same number of unpaired electrons as the target low-spin CSF. 
These orbitals were then localized using the Pipek--Mezey method\cite{Pipek1989} and assigned to each singly-occupied shell to minimize the CSF energy, before subsequent CSF-ROHF optimization.
Since the orbitals in the 30e20o active space from Chan and coworkers are already localized, these calculations were initialized using the same procedure but without the Pipek--Mezey localization step.
All CSF-ROHF solutions were verified by visualizing the orbitals to ensure that they match the expected physical spin coupling.

Benchmark CIPSI calculations were performed using 
Quantum Package 2.0.\cite{garnironQuantumPackage202019}
Starting from the RHF ground state, an initial calculation was converged until the model space contained over $10^6$ determinants. 
The ground-state natural orbitals were then 
computed and the calculation was repeated in this
new basis until convergence was reached or the calculation exceeded the available computational resources.
All calculations were performed without a frozen core approximation.
Best estimates for the target state were computed using a standard linear extrapolation of the variational energy as a function of the second-order perturbation correction.\cite{Sharma2017,Holmes2017,Loos2018b,Burton2024c}

ph-AFQMC calculations were performed using an in-house code and in some cases using the open-source ipie\cite{maloneIpiePythonBasedAuxiliaryField2023} code. 
Further computational details such as ph-AFQMC parameters and twisted ethene geometries can be found in the Supplementary Information.

\section{Acknowledgments}
This work was supported in part by the NOTS cluster operated by Rice University's Center for Research Computing (CRC), and the resources of the Oak Ridge Leadership Computing Facility at the Oak Ridge National Laboratory, which is supported by the Advanced Scientific Computing Research programs in the Office of Science of the U.S. Department of Energy under Contract No. DE-AC05-00OR22725.
H.G.A.B is a Royal Society University Research Fellow (URF\textbackslash R1\textbackslash 241299) at University College London.
J.S. acknowledges support from the Robert A. Welch Foundation (Award Number C-2212) and the American Chemical Society Petroleum Research Fund (\#70053-DNI6).

\section*{Appendices}
\appendix
\section{$L_z$ symmetry in cRHF and in CSFs}\label{appendix:lz}

%%%%%%%%%%%%%%%%%%%%%%%%%%%%%%%%%%%%%%%%%%%%%%%%%%%%%%%%%%%%%%
%\subsection{Orbital angular momentum for linear molecules}
%%%%%%%%%%%%%%%%%%%%%%%%%%%%%%%%%%%%%%%%%%%%%%%%%%%%%%%%%%%%%%
Here we consider the role of the orbital angular momentum for the accuracy of a single CSF compared to cRHF in linear molecules.
Specifically, we consider $\uppi^2$ configurations, which are relevant for \ce{NO-}, \ce{O2}, \ce{NF}, and other linear molecules in the TS12 set.
This derivation follows standard textbook notation for angular momentum coupling.\cite{atkins_mqm} 

Linear systems are invariant to rotations around the internuclear $C_\infty$ axis designated as $z$. 
Therefore, we require eigenstates of the $z$-component of the orbital angular momentum operator, $\hat{L}_z$, of the form
\begin{equation}
\hat{L}_z = - \mathrm{i} \frac{\partial}{\partial \phi}
\end{equation}
where $\phi$ is the angular component in cylindrical polar coordinates.
The angular component of the $\hat{L}_z$ eigenstates are defined as 
\begin{equation}
\psi_m(\phi) = \frac{1}{\sqrt{2\pi}}\mathrm{e}^{\mathrm{i} m \phi}
\end{equation}
where $m$ %in \mathbb{Z}$ 
is an integer.

Eigenstates with the same  $\abs{m}$ share the same eigenvalue of the $\hat{L}_z^2$ operator.
The $m=0$ eigenstate corresponds to the non-degenerate $\Upsigma$ state, which transforms as the totally symmetric irreducible representation. However, for $m \neq 0$, these eigenstates are doubly degenerate and are denoted by the symmetry labels $\Uppi$, $\Updelta$, $\Upphi, \dots$ for $\abs{m} = 1, 2, 3, \dots$ respectively. Due to this degeneracy of the $\pm m$ states, it is always possible to construct real-valued states through linear combinations as
\begin{equation}
\begin{split}
\psi_A &= \frac{1}{\sqrt{2}}(\psi_m(\phi) + \psi_{-m}(\phi)) = \frac{1}{\sqrt{\pi}}\cos(m\phi),
\\
\psi_B &= \frac{\mathrm{i}}{\sqrt{2}}(\psi_m(\phi) - \psi_{-m}(\phi)) = \frac{1}{\sqrt{\pi}}\sin(m\phi).
\end{split}
\end{equation}
For example, these linear combinations for $\abs{m} = 1$ correspond to the $(\uppi_x,\uppi_y)$ orbitals, while for $\abs{m} = 2$ we recover the $(\updelta_{xy},\updelta_{x^2-y^2}$) orbitals. %, as detailed in the character table for $D_{\mathrm{\infty h}}$.
Critically, these real linear combinations are no longer eigenstates of $\hat{L}_z$, but they remain eigenstates of $\hat{L}_z^2$.

We next consider how the many-body basis functions change as we switch between the complex and real angular momentum states. 
In what follows, we use the notation $u$ ($\bar{u}$) to denote the $m=1$ ($-1$) orbitals and $x,y$ for the real orbitals.
Consider the (2e,2o) active space of determinants ordered as $\{ \abs{u \bar{u}}, \abs{u\bar{d}}, \abs{d\bar{u}}, \abs{d\bar{d}} \}$.
The matrix representation of the $\hat{L}_z$ operator in this basis is
\begin{equation}
\bm{L}_z = 
\begin{pmatrix}
        2 & 0 & 0 & 0 \\
        0 & 0 & 0 & 0 \\
        0 & 0 & 0 & 0 \\
        0 & 0 & 0 & -2 
\end{pmatrix}.
\end{equation}
Ensuring that spin symmetry is also conserved, the four eigenstates are $^1\Delta_\text{g}$, $^3\Sigma_\text{g}^-$, and $^1\Sigma_\text{g}^+$, with the functional forms
\begin{subequations}
\begin{align}
^1\Delta_\text{g} &= \abs{u\bar{u}} \text{ and } \abs{d\bar{d}},
\\
^3\Sigma_\text{g}^- &= \frac{1}{\sqrt{2}} (\abs{u\bar{d}} - \abs{d\bar{u}}),
\\
^1\Sigma_\text{g}^+ &= \frac{1}{\sqrt{2}} (\abs{u\bar{d}} + \abs{d\bar{u}}).
\end{align}
\end{subequations}
The two states with $^1\Delta_\text{g}$ symmetry correspond to the $m=2$ and $-2$ states.
%; unlike spherical angular momentum we do not get a $m=0$ or $\pm 1$ component for $\abs{L_z}= 2$.
Under the transformation from complex to real orbitals, we can recover
\begin{subequations}
\begin{align}
^1\Delta_\text{g} &= \frac{1}{2}\Big((\abs{x\bar{x}} - \abs{y\bar{y}}) \pm \mathrm{i} (\abs{x\bar{y}} + \abs{y\bar{x}})\Big),
\label{eq:complexCSF}
\\
^3\Sigma_\text{g}^- &= \frac{1}{\sqrt{2}} (\abs{x\bar{y}} - \abs{y\bar{x}}),
\\
^1\Sigma_\text{g}^+ &= \frac{1}{\sqrt{2}} (\abs{x\bar{x}} + \abs{y\bar{y}}),
\end{align}
\end{subequations}
where any global phase has been removed for clarity.
By a reverse analysis, we see that the open-shell singlet $\Psi_\text{CSF}$ in the real orbital basis can be written in terms of the angular momentum eigenstates as 
\begin{equation}
\Psi_\text{CSF} = \frac{1}{\sqrt{2}}(\abs{x\bar{y}} + \abs{y\bar{x}}) = \frac{\mathrm{i}}{\sqrt{2}}(\abs{u\bar{u}} - \abs{d\bar{d}}).
\label{eq:CSFexpansion}
\end{equation}
Therefore, this CSF is an eigenstate of the total angular momentum $\hat{L}_z^2$ operator, but not an eigenstate of $\hat{L}_z$; in fact we get angular momentum contamination with $\expval*{\hat{L}_z} = 0$.
This is in contrast to the cRHF determinant e.g., $\abs{u\bar{u}}$, which is an eigenstate of both $\hat{L}_z^2$ and $\hat{L}_z$.
As discussed in the main text, we conjecture that this angular momentum contamination can have an adverse effect on the accuracy of the ph-AFQMC dynamics.

% \bibliography{don_references, manuscript, achemso-demo}
\bibliography{ai_format_cited_references}
\end{document}

\maketitle
  The Cholesky decomposition cutoff was set to $10^{-8}$ unless mentioned otherwise. Additional details are found below.

\subsection{HHeH in cc-pVDZ basis}
%All ph-AFQMC calculations were run using an in-house C++/HIP code.

ph-AFQMC calculations used 2048 walkers and 4000 blocks, where each block consists of 20 steps with timestep $\Delta\tau=0.005\text{ Ha}^{-1}$. Walkers were initialized by sampling from the multi-Slater determinant trial (for this CSF-ROHF state with $+-$ coupling, that would correspond to an even mixture of the 2 determinants).

\subsection{TS12 in aug-cc-pVTZ and aug-cc-pVQZ bases}

The TS12 geometries were taken from Ref. \citenum{leeTwoSinglereferenceApproaches2019}, and the ph-AFQMC energies with cRHF and UHF trials from Ref. \citenum{leeUtilizingEssentialSymmetry2020}. 
%The aug-cc-pvtz and aug-cc-pvqz CSF-ROHF trial calculations used in CBS were run using a group-developed code. 
%A cholesky cutoff of $10^{-8}$ was used for this dataset. 
We used 8192 walkers and 5000 blocks for the ph-AFQMC calculations, where each block consists of 20 steps with timestep $\Delta\tau=0.005\text{ Ha}^{-1}$. Walkers were initialized by sampling from the trial.
The complete basis set (CBS) extrapolation was performed using the protocol of Helgaker \citenum{helgaker1997basis}, %\footnote{Helgaker et al. JCP (1997) 106 23 9639-9646} equation 
treating aug-cc-pVQZ RHF as the CBS HF value.

\begin{table}[h]
\centering
\begin{tabular}{lcr}
%\toprule
\hline\hline
System & Active space & CASSCF energy [Ha] \\
%\midrule
\hline 
C & 2e3o & -37.633 243 \\
NF & 8e6o & -153.819 645 \\
NH & 4e4o & -54.930 521 \\
NO$^-$ & 8e6o & -129.297 659 \\
O$_2$ & 8e6o & -149.721 164 \\
O & 4e3o & -74.726 825 \\
PF & 8e6o & -440.203 452 \\
PH & 4e4o & -341.264 036 \\
S$_2$ & 8e6o & -795.104 079 \\
S & 4e3o & -397.453 798 \\
Si & 2e3o & -288.818 208 \\
SO & 8e6o & -472.421 823 \\
\hline\hline
\end{tabular}
\caption{CASSCF active spaces and associated CASSCF energies used in ph-AFQMC calculations.}
\label{table:AS}
\end{table}

The CASSCF active spaces used to construct the 2-determinant CASSCF-orbital along with the CASSCF + $L_z$ trial wavefunctions are shown in \autoref{table:AS} using the aug-cc-pVTZ basis.
ph-AFQMC calculations using the ($L_z$ adapted) CSF-ROHF and ($L_z$ adapted) 2-determinant CASSCF trials used ipie, which can use complex CI coefficients.  For a maximally direct comparison to data obtained using our group's in-house code, we used the same Cholesky cutoff of $10^{-8}$ and time-step, with 8192 walkers and 5000 blocks.  In the interests of data reproducibility, a full ipie python run script and all associated .chk files are included in the supplementary information. Walkers were initialized from the RHF determinant with a small  ($10^{-5}$) amount of additional random noise necessary for calculation stability.  %As a consistency check, the biggest difference in a CSF-ROHF trial calculation between the two codes was $0.4\text{ kcal/mol}$ for O$_2$.

In the reblocking procedure, the first 2000 blocks were discarded and the reblocking iteration that led to the lowest stochastic error was selected (retaining at least 20\% of the full dataset $=1000$ blocks). Due to large spikes in the ipie AFQMC trajectories an extra `filter' step was used with the following thresholds:

\begin{table}[h]
\centering
%\resizebox{\textwidth}{!}{
\begin{tabular}{lccccc}
\hline\hline 
 & Reblocking filter & \multicolumn{4}{c}{Points removed} \\  \cline{3-6} 
System & threshold / Ha & 2-det-CAS & 2-det-CAS$+L_z$ &  CSF & CSF$+L_z$ \\
\hline 
C & 0.01 & 6 & 1 & 0 & 6 \\
NF & 0.02 & 3 & 9 & 2 & 12 \\
NH & 0.01 & 1 & 0 & 2 & 4 \\
NO$^-$ & 0.03 & 0 & 4 & 3 & 4 \\
O & 0.01 & 0 & 4 & 1 & 2 \\
O$_2$ & 0.04 & 0 & 3 & 0 & 8 \\
PF & 0.02 & 1 & 1 & 1 & 3 \\
PH & 0.01 & 1 & 0 & 1 & 2 \\
S & 0.01 & 0 & 2 & 0 & 3 \\
S$_2$ & 0.02 & 0 & 2 & 0 & 3 \\
Si & 0.005 & 3 & 32 & 3 & 33 \\
SO & 0.02 & 4 & 15 & 3 & 23 \\
\hline  \hline
\end{tabular}
%}
\end{table}

\subsection{Twisted ethene in aug-cc-pVDZ basis}
The XYZ coordinates of the 2 geometries studied (in Angstrom):

\vspace{0.5em}
\noindent $\phi_{\text{H-C-C-H}} = 82.5^\circ$
\begin{verbatim}
C   -0.66690500   0.00000000   0.00000000
H   -1.22952232   0.92228948   0.00000000
H   -1.22952232  -0.92228948   0.00000000
C    0.66690500   0.00000000   0.00000000
H    1.22952232   0.12038293   0.91439916
H    1.22952232  -0.12038293  -0.91439916
\end{verbatim}

\noindent $\phi_{\text{H-C-C-H}} = 90^\circ$
\begin{verbatim}
C   -0.66690500   0.00000000   0.00000000
H   -1.22952232   0.92228948   0.00000000
H   -1.22952232  -0.92228948   0.00000000
C    0.66690500   0.00000000   0.00000000
H    1.22952232   0.00000000   0.92228948
H    1.22952232  -0.00000000  -0.92228948
\end{verbatim}

%The associated HF and coupled cluster energies were obtained with PySCF.

ph-AFQMC calculations %were run using a group-developed C++/HIP code. A cholesky cutoff of $10^{-8}$ was used for this dataset. Imaginary time propagation was performed 
used 32768 walkers and 5000 blocks. %, each block consisting of 20 microblocks, each with $\Delta T=0.005\text{ Ha}^{-1}$. 
For the CSF-ROHF trial calculations walkers were initialized by sampling from the multi-Slater determinant trial; for the UHF trial calculations they were initialized as UHF. 

\subsection{[2Fe--2S]$^{2+}$ active space}
We have used the publicly available integral files for this active space from \href{https://github.com/zhendongli2008/Active-space-model-for-Iron-Sulfur-Clusters}{github}.
For the CSF we used a coupling pattern of $+++++-----$ (corresponding to 252 determinants).
ph-AFQMC calculations used 2048 walkers and 5000 blocks, each block consisting of 20 microblocks, each with $\Delta T=0.005\text{ Ha}^{-1}$. 
For the CSF-ROHF trial walkers were initialized by sampling from the multi-Slater determinant trial.
As mentioned in the main text a CI coefficient cutoff of 99.95\% was used for the 10e10o CASSCF wavefunction corresponding to 8943 determinants.

\subsection{[2Fe--2S]$^{2+}$ full space in cc-pVDZ basis}

For the CSF we used the same coupling pattern as in the active space, which corresponded to 252 determinants.
In the ph-AFQMC calculations, 44 core orbitals were frozen corresponding to the $6 \times \text{S}(1s^2 2s^2) + 2 \times \text{Fe}(1s^2 2s^2 2p^6)$.

A slightly looser Cholesky cutoff of $10^{-6}$ was used for this dataset due to the larger system size.  We used 8192 walkers and 5000 blocks, where each block consists of 20 steps with timestep $\Delta\tau=0.005\text{ Ha}^{-1}$. For calculations with the CASSCF or CSF-ROHF trial, walkers were initialized by sampling from the multi-Slater determinant trial. For the CASSCF trial, natural orbitals were used.

\bibliography{ai_format_cited_references}